\documentclass[preprint, aps, pra]{revtex4}	
\usepackage{latexsym,amssymb,makeidx}

\usepackage{epsfig, color, ulem}
\usepackage{graphicx}
\usepackage{dcolumn}
\usepackage{bm}
\usepackage{amsmath}
\usepackage{setspace}
\usepackage{lineno}

\begin{document}

\def \rev#1{\textcolor{black}{#1}}

\title{Unified wave field retrieval and imaging  method for inhomogeneous non-reciprocal media}
\author{Kees Wapenaar and Christian Reinicke}
\affiliation{Department of Geoscience and Engineering,\\ Delft University of Technology,\\ Stevinweg 1,\\ 2628 CN Delft,\\ The Netherlands}



\begin{spacing}{1.3}

\def\a{\rm (c)}
\def\ap{+\a}
\def\am{-\a}
\def\bN{{\bf N}}
\def\bK{{\bf K}}
\def\bJ{{\bf J}}
\def\barrho{\rho^o\!\!}
\def\bareps{\varepsilon^o\!\!}
\def\bA{{{\mbox{\boldmath ${\cal A}$}}}}
\def\bAp{\,\,\,\tilde{\!\!\!\bA}}
\def\bL{{{\mbox{\boldmath ${\cal L}$}}}}
\def\bH{{{\mbox{\boldmath ${\cal H}$}}}}
\def\bAt{\bAp}
\def\bLt{\,\tilde{\!\bL}}
\def\bHt{\tilde\bH}
\def\bq{{\bf q}}
\def\bd{{\bf d}}
\def\bp{{\bf p}}
\def\bs{{\bf s}}
\def\u{U}
\def\m{m}
\def\f{P}
\def\sigmaa{p}
\def\sigmaaa{\sigmaa}
\def\ddelta{\chi}
\def\s{\vartheta}
\def\h{\s_{33}}
\def\ss{d}
\def\bb{b}
\def\K{K}
\def\F{F}
\def\p{s_1}
\def\q{s_3}
\def\bx{{\bf x}}
\def\bxa{{\bf x}_A}
\def\bxr{{\bf x}_S}
\def\bxh{x_1}
\def\bxha{x_{1,A}}
\def\bxhr{x_{1,S}}
\def\setR{\mathbb{R}}
\def\half{\begin{matrix}\frac{1}{2}\end{matrix}}
\def\third{\begin{matrix}\frac{1}{3}\end{matrix}}
\def\quart{\begin{matrix}\frac{1}{4}\end{matrix}}
\def\ninth{\begin{matrix}\frac{1}{9}\end{matrix}}
\def\pa {{\bf p}_A}
\def\pb {{\bf p}_B}
\def\R {{\bf R}}
\def\T {{\bf T}}
\def\G {{\bf G}}
\def\M {{\bf C}}
\def\I {{\bf I}}
\def\O {{\bf O}}
\def\r{{\bf r}^\cap}
\def\too{\to}
\def\setD{\mathbb{D}}
\def\setdD{{\partial\setD}}
\def\xa{x_{3,0}}
\def\xb{x_{3,A}}
\def\i{i}
\def\dxx{{\rm d}x_1}
\def\ddx{{\rm d}\bx}
\def\dx{\ddx}
\def\dpp{{\rm d}\p }
\def\dt{{\rm d}t'}
\def\etaa{S^{(2)}}
\def\thetaa{S^{(1)}}

\begin{abstract}
Acoustic imaging methods often ignore multiple scattering. This leads to false images in cases where multiple scattering is strong.
Marchenko imaging has recently been introduced as a data-driven way to deal with internal multiple scattering. 

Given the increasing interest in non-reciprocal materials, both for acoustic and electromagnetic applications, we propose to modify the Marchenko method for 
imaging such materials.
We formulate a unified wave equation for non-reciprocal materials, exploiting the similarity between acoustic and electromagnetic wave phenomena.
This unified wave equation forms the basis for deriving reciprocity theorems that interrelate wave fields in a non-reciprocal medium and its  \rev{complementary version}. 
Next, we reformulate these theorems for downgoing and upgoing wave fields. 
From these decomposed reciprocity theorems we derive representations of the Green's function inside the non-reciprocal medium, in terms of the reflection response
at the surface and focusing functions inside the medium and its  \rev{complementary version}. 
These representations form the basis for deriving a modified version of the Marchenko method  \rev{to retrieve the wave field inside a}
non-reciprocal medium  \rev{and to form an image, free from artefacts related to multiple scattering}. 
We illustrate the proposed method at the hand of the numerically modeled reflection response of a horizontally layered medium.
\end{abstract}

\maketitle

\section{Introduction}

Acoustic imaging methods are traditionally based on the single-scattering assumption \cite{Claerbout71GEO, Stolt78GEO, Berkhout79GP, Williams80PRL, Devaney82UI, 
Bleistein82GEO, Maynard85JASA, Langenberg1986NDT, McMechan83GP, Esmersoy88GEO, Oristaglio89IP, Norton92JASA, Wu2004JASA, Lindsey2004AJSS, Etgen2009GEO}.
Multiply scattered waves are not properly handled by these methods and may lead to false images overlaying the desired primary image. 
Several approaches have been developed that account for multiple scattering. 
For the sake of the discussion it is important to distinguish between different classes of multiply scattered waves.
Waves that have scattered at least once at the surface of the medium are called surface-related multiples. 
This type of multiple scattering is particularly severe in exploration geophysics. However, because the scattering boundary is known, this class of multiples is relatively easily dealt with.
Successful methods have been developed to suppress surface-related multiples prior to
imaging \cite{Verschuur92GEO, Carvalho92SEG, Borselen96GEO, Biersteker2001SEG, Pica2005TLE, Dragoset2010GEO}.
Waves that scatter several times inside the medium before being recorded at the surface are called internal multiples. 
Internal multiple scattering may occur at heterogeneities at many scales. 
We may distinguish between deterministic scattering at well-separated scatterers, giving rise to long period multiples, and diffuse scattering in stochastic media.
Of course this distinction is not always sharp. In this paper we only consider the first type of internal multiple scattering, which typically occurs in layered media
(which, in general, may have curved interfaces and varying parameters in the layers).
Several imaging approaches that account for deterministic internal multiples are currently under development, such as the inverse scattering series 
approach \cite{Weglein97GEO, Kroode2002WM, Weglein2003IP}, full wave field migration \cite{Berkhout2014GP, Davydenko2017GP}, and Marchenko imaging.
The latter approach builds on a 1D autofocusing procedure \cite{Rose2001PRA, Rose2002IP, Broggini2012EJP}, which has been generalised for 2D and 3D inhomogeneous 
media \cite{Wapenaar2012GJI, Wapenaar2014JASA, Broggini2014GEO, Behura2014GEO, Meles2015GEO, Neut2015GJI, Neut2016GEO, Thorbecke2017GEO, 
Neut2017GEO, Singh2017GEO, Mildner2017GEO, Elison2018GJI}.
This methodology  \rev{retrieves the wave fields inside a medium, including all internal multiples,
in a data-driven way. Such wave fields could be used, for example, to monitor changes of the material over time. 
Moreover, in a next step these wave fields can be used to form an image of the material, in which artefacts due to the internal multiples are suppressed.} 
Promising results have been obtained with 
geophysical \cite{Ravasi2016GJI, Ravasi2017GEO, Staring2018GEO, Brackenhoff2019SE, Wapenaar2018SR} and ultrasonic data  \rev{ \cite{Wapenaar2018SR, Cui2018PRA}}.

To date, the application of the Marchenko  \rev{method} has been restricted to reciprocal media. With the increasing interest in non-reciprocal materials,
both in electromagnetics \cite{Willis2011RS, He2011PRB, Ardakani2014JOSA} 
and in acoustics  \rev{and elastodynamics \cite{Willis2012CRM, Norris2012RS, Gu2016SR, Trainiti2016NJP,  Nassar2017RS, Nassar2017JMPS, Attarzadeh2018JSV}}, 
it is opportune to modify the Marchenko method for  non-reciprocal media.
We start with a brief review of the wave equation for non-reciprocal media. By restricting this to scalar waves in a 2D plane, it is possible to capture
different wave phenomena by a unified wave equation.
Next, we formulate reciprocity theorems for waves in a non-reciprocal medium and its  \rev{complementary version (the complementary medium will be defined later)}. 
From these reciprocity theorems we derive Green's function representations, which form the basis 
for the Marchenko method in non-reciprocal media. We illustrate the new method with a numerical example, 
showing that it has the potential to accurately  \rev{retrieve the wave fields inside} a non-reciprocal medium  \rev{and to image this medium}, without
false images related to multiply scattered waves.

\section{Unified wave equation for non-reciprocal media}

Consider the following unified equations  \rev{in the low-frequency limit} for 2D wave propagation in the $(x_1,x_3)$-plane in inhomogeneous, lossless, anisotropic, non-reciprocal media
\begin{eqnarray}
&&\alpha \partial_tP +   (\partial_r+ \gamma_r\partial_t)Q_r =B,\label{eq15agt1}\\
&&(\partial_r+ \gamma_r\partial_t)P+\beta_{rs}\partial_tQ_s =C_r.\label{eq16agt1}
\end{eqnarray}
These equations hold for transverse-electric (TE), transverse-magnetic (TM), horizontally-polarised shear (SH) and acoustic (AC) waves.
They are formulated in the space-time $(\bx,t)$ domain, with  $\bx=(x_1,x_3)$.
Operator $\partial_r$ stands for differentiation in the $x_r$ direction. 
Lower-case subscripts $r$ and $s$ take the values 1 and 3 only; Einstein's summation convention applies for repeated subscripts. Operator $\partial_t$ stands for temporal differentiation.
 \rev{The wave field quantities ($P=P(\bx,t)$ and $Q_r=Q_r(\bx,t)$) and source quantities  ($B=B(\bx,t)$ and $C_r=C_r(\bx,t)$) are macroscopic quantities. These are 
often denoted as $\langle P\rangle$ etc. \cite{Willis2011RS}, but for notational convenience we will not use the brackets. The medium parameters
($\alpha=\alpha(\bx)$, $\beta_{rs}=\beta_{rs}(\bx)$ and $\gamma_r=\gamma_r(\bx)$) are effective parameters. In general they are anisotropic at macro scale (with $\beta_{rs}=\beta_{sr}$), 
even when they are isotropic at micro scale.
Wave field quantities, source quantities and medium parameters are specified for the different wave phenomena in Table 1.
For TE and TM waves, the macroscopic wave field quantities are $E$  (electric field strength) and $H$ (magnetic field strength), 
the macroscopic source functions are $J^{\rm e}$  (external electric current density) and $J^{\rm m}$  (external magnetic current density),
and the effective medium parameters are $\bareps\,$ (permittivity), $\mu$ (permeability) and $\xi$ (coupling parameter).
For SH and AC waves,  the macroscopic wave field quantities are $v$ (particle velocity), $\tau$ (stress) and $\sigmaaa$ (acoustic pressure),
the macroscopic source functions are $\F$ (external force density), $h$ (external deformation-rate density) and $q$ (volume injection-rate density),
and the effective medium parameters are $\barrho\,$ (mass density), $s$ (compliance), $\kappa$ (compressibility) and $\xi$ (coupling parameter).
For further details we refer to Appendix \ref{AppA}.}

\begin{center}
{{\noindent \it Table 1: Quantities in unified equations (\ref{eq15agt1}) and (\ref{eq16agt1}).}
\begin{tabular}{||l|c|c|c|c|c|c|c|c|c|c|c|c|c|c|c||}
\hline\hline
& $P$ & $Q_1$ & $Q_3$ & $\alpha$ &$\beta_{11}$ &$\beta_{31}$  & $\beta_{33}$ & $\gamma_1$ & $\gamma_3$ & $B$ & $C_1$ & $C_3$ \\
\hline
TE & $E_2$ &$H_3$ & $-H_1$ & ${\bareps}_{22}$ &$\mu_{33}$ &$-\mu_{31}$ & $\mu_{11}$ &$\xi_{23}$ & $-\xi_{21}$  & $-J_2^{\rm e}$ & $-J_3^{\rm m}$& $J_1^{\rm m}$  \\
\hline
TM & $H_2$ &$-E_3$ & $E_1$ & $\mu_{22}$ &${\bareps}_{33}$ &$-{\bareps}_{31}$ & ${\bareps}_{11}$ &$-\xi_{32}$&  $\xi_{12}$  & $-J_2^{\rm m}$ & $J_3^{\rm e}$& $-J_1^{\rm e}$ \\
\hline
SH  & $v_2$ &$-\tau_{21}$ & $-\tau_{23}$  & ${\barrho}_{22}$ &$4s_{1221}$ &$4s_{1223}$ &  $4s_{3223}$&$2\xi_{221}$& $2\xi_{223}$   &$\F_2$ & $2h_{21}$ & $2h_{23}$  \\
\hline
AC  & $\sigmaaa$ &$v_1$ & $v_3$  & $\kappa$ &${\barrho}_{11}$ &${\barrho}_{31}$ &  ${\barrho}_{33}$&$\xi_1$& $\xi_3$   &$q$ & $\F_1$ & $\F_3$  \\
\hline
\hline
\end{tabular}
}
\end{center}
\mbox{}\\

By eliminating $Q_r$ from equations (\ref{eq15agt1}) and (\ref{eq16agt1}) we obtain a scalar wave equation for field quantity $P$, according to
\begin{eqnarray}
&& (\partial_r+\gamma_r\partial_t) \s_{rs}(\partial_s+\gamma_s\partial_t)P-\alpha\partial_t^2P=(\partial_r+\gamma_r\partial_t)\s_{rs}C_s-\partial_tB,\label{eq16aghht}
\end{eqnarray}
see Appendix \ref{AppA} for the derivation. Here $\s_{rs}$ is the inverse of $\beta_{rs}$. 
 Compare equation (\ref{eq16aghht}) with the common  \rev{scalar} wave equation for waves in isotropic reciprocal media 
\begin{eqnarray}
&&\partial_r\frac{1}{\beta}\partial_rP-\alpha\partial_t^2P=\partial_r\frac{1}{\beta}C_r-\partial_tB.\label{eq16aghhtir}
\end{eqnarray}
In equation (\ref{eq16aghht}), $\partial_r+\gamma_r\partial_t$ replaces $\partial_r$, with $\gamma_r$ being responsible for the non-reciprocal behaviour. Moreover, 
 $\s_{rs}$ replaces $1/\beta$, thus accounting for anisotropy of the effective non-reciprocal medium.

 \rev{To illustrate the physical meaning of the parameter $\gamma_r$, we consider the 1D version of equation (\ref{eq16aghht}) for a homogeneous, isotropic, source-free medium, i.e.
\begin{eqnarray}
&& (\partial_1+\gamma\partial_t)(\partial_1+\gamma\partial_t)P-\alpha\beta\partial_t^2P=0.
\end{eqnarray}
Its solution reads
\begin{eqnarray}
&& P^\pm(x_1,t)=S\Bigl(t\mp \frac{x_1}{c}(1\pm\gamma c)\Bigr),
\end{eqnarray}
with  $S(t)$ being an arbitrary time-dependent function and $c=(\alpha\beta)^{-1/2}$ the propagation velocity of the corresponding reciprocal medium. 
Note that $P^+(x_1,t)$ propagates in the positive $x_1$-direction with 
slowness $(1+\gamma c)/c$, whereas $P^-(x_1,t)$ propagates in the negative $x_1$-direction with slowness $(1-\gamma c)/c$. 
Hence, $\gamma$ determines the asymmetry of the slownesses in opposite directions.
Throughout this paper we assume that $|\gamma_r|$ is smaller than the lowest inverse propagation velocity of the corresponding reciprocal anisotropic medium.
}

\section{Reciprocity theorems for a non-reciprocal medium and its complementary version}

We derive reciprocity theorems in the space-frequency $(\bx,\omega)$-domain for wave fields in a non-reciprocal medium and its  \rev{complementary version}.
To this end, we define the temporal Fourier transform of a space- and time-dependent function $\f(\bx,t)$  as
\begin{equation}\label{eqFT}
\f(\bx,\omega)=\int_{-\infty}^\infty\f(\bx,t) \exp(\i\omega t){\rm d}t,
\end{equation}
where $\omega$ is the angular frequency and  $\i$ the imaginary unit. For notational convenience we use the same symbol 
for quantities in the time domain and in the frequency domain.
We use equation (\ref{eqFT}) to transform equations (\ref{eq15agt1}) and (\ref{eq16agt1}) to the space-frequency domain. 
The temporal differential operators $\partial_t$ are thus replaced by $-\i\omega$, hence
\begin{eqnarray}
&&-\i\omega\alpha P + (  \partial_r-\i\omega \gamma_r)Q_r =B,\label{eq15agf}\\
&&(\partial_r-\i\omega \gamma_r)P-\i\omega\beta_{rs}Q_s =C_r,\label{eq16agf}
\end{eqnarray}
with $P=P(\bx,\omega)$, $Q_r=Q_r(\bx,\omega)$, $B=B(\bx,\omega)$ and $C_r=C_r(\bx,\omega)$.
A reciprocity theorem formulates a mathematical relation between two independent states  \cite{Fokkema93Book, Hoop95Book, Achenbach2003Book}. 
We indicate the wave fields, sources and medium parameters  in the two states by subscripts $A$ and $B$. 
Consider the quantity
\begin{equation}
\partial_r(P_AQ_{r,B}-Q_{r,A}P_B).
\end{equation}
Applying the product rule for differentiation, using equations (\ref{eq15agf}) and (\ref{eq16agf}) for states $A$ and $B$, using $\beta_{sr}=\beta_{rs}$  \rev{ \cite{Nassar2017JMPS, Kong72IEEE, Birss67PM}},
integrating the result over domain $\setD$ enclosed by boundary $\setdD$ with outward pointing normal vector ${\bf n}=(n_1,n_3)$
 and applying the theorem of Gauss, we obtain
\begin{eqnarray}
&&\oint_\setdD(P_AQ_{r,B}-Q_{r,A}P_B)n_r{\rm d}\bx=\\
&&\i\omega\int_\setD\Bigl((\alpha_B-\alpha_A)P_AP_B-(\beta_{rs,B}-\beta_{rs,A})Q_{r,A}Q_{s,B}
\Bigr)\ddx\nonumber\\
&&+\i\omega\int_\setD(\gamma_{r,B}+\gamma_{r,A})(P_AQ_{r,B}-Q_{r,A}P_B)\ddx\nonumber\\
&&+\int_\setD(C_{r,A}Q_{r,B}-Q_{r,A}C_{r,B}+P_AB_B-B_AP_B)\ddx.\nonumber
\end{eqnarray}
This is the general reciprocity theorem of the convolution type.
When the medium parameters $\alpha$, $\beta_{rs}$ and $\gamma_r$ are identical in both states, then the first integral on the right-hand side vanishes, but the second
integral, containing $\gamma_r$, does not vanish. 
When we choose  $\gamma_{r,A}=-\gamma_{r,B}=-\gamma_r$, then the second integral also vanishes. 
For this situation we call state $B$, with parameters $\alpha$, $\beta_{rs}$ and $\gamma_r$, the actual state, and state $A$, with parameters
$\alpha$, $\beta_{rs}$ and $-\gamma_r$, the  \rev{complementary state \cite{Kong72IEEE, Lindell95JEVA} (also known as the Lorentz-adjoint state \cite{Altman91Book}).}
We indicate the  \rev{complementary} state by a superscript $\a$. Hence
\begin{eqnarray}\label{reccon}
&&\oint_\setdD(P_A^{\a}Q_{r,B}-Q_{r,A}^{\a}P_B)n_r{\rm d}\bx=\\
&&\int_\setD(C_{r,A}^{\a}Q_{r,B}-Q_{r,A}^{\a}C_{r,B}+P_A^{\a}B_B-B_A^{\a}P_B)\ddx.\nonumber
\end{eqnarray}
This reciprocity theorem will play a role in the derivation of Green's function representations for the Marchenko method for non-reciprocal media (section \ref{Marchenko}). Here we 
use it to derive a relation between Green's functions in states $A$ and $B$. For the  \rev{complementary} state $A$ we choose a unit monopole point source at $\bx_S$ in $\setD$, hence
$B^{\a}_A(\bx,\omega)=\delta(\bx-\bx_S)$, where $\delta(\bx)$ is the Dirac delta function. The response to this point source is the Green's function in state $A$, hence
 $P^{\a}_A(\bx,\omega)=G^{\a}(\bx,\bx_S,\omega)$. Similarly, for state $B$ we choose a unit monopole point source at $\bx_R$ in $\setD$, hence
$B_B(\bx,\omega)=\delta(\bx-\bx_R)$ and $P_B(\bx,\omega)=G(\bx,\bx_R,\omega)$. We substitute these expressions into equation (\ref{reccon}) and set the other source quantities, 
$C_{r,A}^{\a}$ and $C_{r,B}$, to zero. Further, we assume that Neumann or Dirichlet boundary conditions apply at $\setdD$, or that the medium at and outside $\setdD$ is homogeneous
and reciprocal. In each of  these cases the boundary integral vanishes. We thus obtain \cite{Slob2009PIER, Willis2012CRM}
\begin{equation}\label{eq11}
G^{\a}(\bx_R,\bx_S,\omega)=G(\bx_S,\bx_R,\omega).
\end{equation}
The left-hand side is the response to a source at $\bx_S$ in the  \rev{complementary} medium (with parameter $-\gamma_r$), observed by a receiver at $\bx_R$.
The right-hand side is the response to a source at $\bx_R$ in the actual medium (with parameter $\gamma_r$), observed by a receiver at $\bx_S$. 
Note the analogy with the flow-reversal theorem for waves in flowing media \cite{Lyamshev61DAN, Godin97WM, Wapenaar2004ACME}.

\noindent
Next, we consider the quantity
\begin{equation}
\partial_r(P_A^*Q_{r,B}+Q_{r,A}^*P_B).
\end{equation}
Superscript $*$ denotes complex  conjugation.
Following the same steps as before, we obtain
\begin{eqnarray}
&&\oint_\setdD(P_A^*Q_{r,B}+Q_{r,A}^*P_B)n_r{\rm d}\bx=\\
&&\i\omega\int_\setD\Bigl((\alpha_B-\alpha_A)P_A^*P_B+(\beta_{rs,B}-\beta_{rs,A})Q_{r,A}^*Q_{s,B}
\Bigr)\ddx\nonumber\\
&&+\i\omega\int_\setD(\gamma_{r,B}-\gamma_{r,A})(P_A^*Q_{r,B}+Q_{r,A}^*P_B)\ddx\nonumber\\
&&+\int_\setD(C_{r,A}^*Q_{r,B}+Q_{r,A}^*C_{r,B}+P_A^*B_B+B_A^*P_B)\ddx.\nonumber
\end{eqnarray}
This is the general reciprocity theorem of the correlation type.
When the medium parameters $\alpha$, $\beta_{rs}$ and $\gamma_r$ are identical in both states, then the first and second integral on the right-hand side vanish.
Hence
\begin{eqnarray}\label{reccor}
&&\oint_\setdD(P_A^*Q_{r,B}+Q_{r,A}^*P_B)n_r{\rm d}\bx=\\
&&\int_\setD(C_{r,A}^*Q_{r,B}+Q_{r,A}^*C_{r,B}+P_A^*B_B+B_A^*P_B)\ddx.\nonumber
\end{eqnarray}
Also this reciprocity theorem will play a role in the derivation of Green's function representations for the Marchenko method for non-reciprocal media.

\begin{figure}
\vspace{0cm}
\centerline{\epsfysize=8 cm \epsfbox{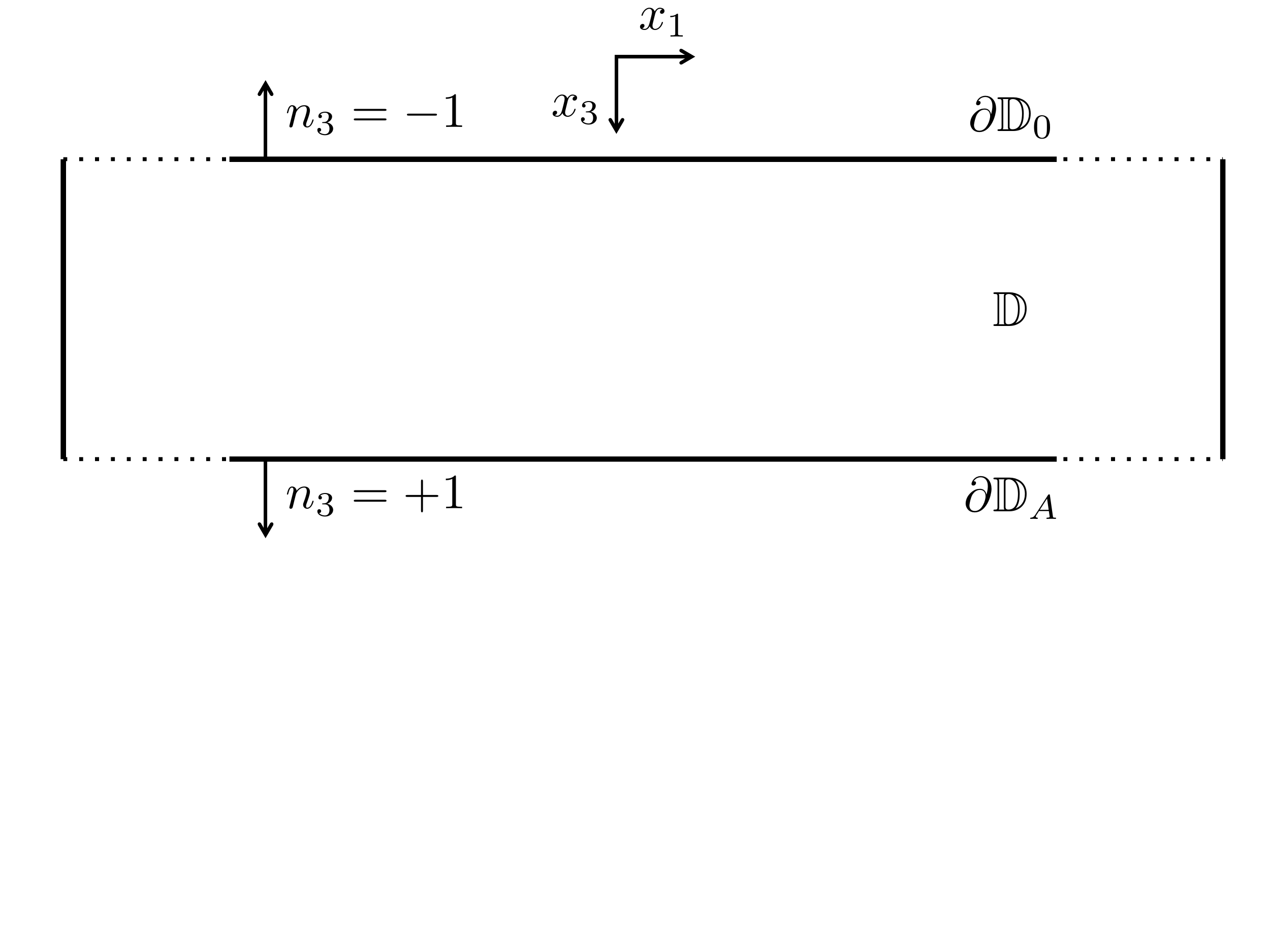}}
\vspace{-2.8cm}
\caption{\footnotesize Modified configuration for the reciprocity theorems. 
}\label{Fig1}
\end{figure}

\section{Green's function representations for the Marchenko method}\label{Marchenko}

We use the reciprocity theorems of the convolution and correlation type (equations (\ref{reccon}) and (\ref{reccor})) to derive Green's function 
representations for the Marchenko method for non-reciprocal media. 
The derivation is similar to that for reciprocal media \cite{Wapenaar2014JASA}; here we emphasise the differences.
We consider a spatial domain $\setD$, enclosed by two infinite horizontal boundaries $\setdD_0$ and $\setdD_A$ (with $\setdD_A$ below $\setdD_0$), 
and two finite vertical side boundaries (at $x_1\to\pm\infty$), see Figure \ref{Fig1}.  The positive $x_3$-axis points downward.
The normal vectors at $\setdD_0$ and $\setdD_A$ are ${\bf n}=(0,-1)$ and ${\bf n}=(0,1)$, respectively. 
The boundary integrals in equations (\ref{reccon}) and (\ref{reccor}) along the vertical side boundaries vanish \cite{Wapenaar89Book}.
Assuming there are no sources in $\setD$ in both states, the reciprocity theorems thus simplify to
\begin{eqnarray}\label{eq19g}
&&\int_{\setdD_0}(P_A^{\a}Q_{3,B}-Q_{3,A}^{\a}P_B)\dx=\int_{\setdD_A}(P_A^{\a}Q_{3,B}-Q_{3,A}^{\a}P_B)\dx
\end{eqnarray}
and
\begin{eqnarray}\label{eq20g}
&&\int_{\setdD_0}(P_A^*Q_{3,B}+Q_{3,A}^*P_B)\dx=\int_{\setdD_A}(P_A^*Q_{3,B}+Q_{3,A}^*P_B)\dx.
\end{eqnarray}
For the derivation of the representations for the Marchenko method it is convenient to decompose the wave field quantities in 
these theorems into downgoing and upgoing fields in both states.
Consider the following relations 
\begin{equation}\label{eqcomp}
\bq=\bL\bp,\quad \bp=\bL^{-1}\bq,
\end{equation}
with wave vectors $\bq=\bq(\bx,\omega)$ and $\bp=\bp(\bx,\omega)$ defined as
\begin{eqnarray}\label{eq23ag}
\bq =\begin{pmatrix} P \\ Q_3 \end{pmatrix}, 
\quad \bp =\begin{pmatrix} \u^+ \\ \u^- \end{pmatrix}.
\end{eqnarray}
Here $\u^+=\u^+(\bx,\omega)$ and $\u^-=\u^-(\bx,\omega)$ are downgoing and upgoing  \rev{flux-normalized}  wave fields, respectively. Operator $\bL=\bL(\bx,\omega)$ in equation (\ref{eqcomp})
is a pseudo-differential operator that composes the total wave field from its downgoing and upgoing 
constituents \cite{Corones83JASA, Fishman87JASA, Wapenaar89Book, Fishman93RS, Hoop92PHD, Hoop96JMP, Wapenaar96JASA, Haines96JMP, Fishman2000JMP}. 
Its inverse decomposes the total wave field into downgoing and upgoing fields.
For inhomogeneous isotropic reciprocal media, the theory for this operator is well developed.
For anisotropic non-reciprocal media, we restrict the application of this operator to the laterally invariant situation. In Appendix \ref{AppB} we use equations 
(\ref{eqcomp}) and (\ref{eq23ag}) at boundaries $\setdD_0$
and $\setdD_A$ to recast reciprocity theorems (\ref{eq19g}) and (\ref{eq20g}) as follows
\begin{eqnarray}
&&\int_{\setdD_0}\bigl(\u_A^{\ap}\u_B^- - \u_A^{\am}\u_B^+\bigr)\dx=
\int_{\setdD_A}\bigl(\u_A^{\ap}\u_B^- - \u_A^{\am}\u_B^+\bigr)\dx\label{eq142}
\end{eqnarray}
and
\begin{eqnarray}
&&\int_{\setdD_0}\bigl(\u_A^{+*}\u_B^+ - \u_A^{-*}\u_B^-\bigr)\dx=
\int_{\setdD_A}\bigl(\u_A^{+*}\u_B^+ - \u_A^{-*}\u_B^-\bigr)\dx.\label{eq143}
\end{eqnarray}
Equation (\ref{eq142}) is exact, whereas in equation (\ref{eq143}) evanescent waves are neglected at boundaries $\setdD_0$ and $\setdD_A$.
Note that the assumption of lateral invariance only applies to boundaries $\setdD_0$ and $\setdD_A$; the remainder of the medium (in- and outside $\setD$) 
may be arbitrary inhomogeneous.

In the following we define $\setdD_0$ (at $x_3=x_{3,0}$) as the upper boundary  of an inhomogeneous, anisotropic, non-reciprocal, lossless medium.
Furthermore, we define $\setdD_A$ (at $x_3=x_{3,A}$, with $x_{3,A}>x_{3,0}$) as an arbitrary boundary inside the medium.
We assume that the medium above $\setdD_0$ is homogeneous. 
For state $B$ we consider a unit source for downgoing waves at $\bxr=(x_{1,S},x_{3,S})$, just above $\setdD_0$ (hence, $x_{3,S}=x_{3,0}-\epsilon$, with $\epsilon\to 0$).
The response to this unit source at any observation point $\bx$ is given by $\u_B^\pm(\bx,\omega)=G^\pm(\bx,\bxr,\omega)$, where $G^+$ and $G^-$ denote the 
downgoing and upgoing components of the Green's function.
For $\bx$ at $\setdD_0$, i.e., just below the source,  we have $\u_B^+(\bx,\omega)=G^+(\bx,\bxr,\omega)=\delta(\bxh-\bxhr)$ and 
$\u_B^-(\bx,\omega)=G^-(\bx,\bxr,\omega)=R(\bx,\bxr,\omega)$, with $R(\bx,\bxr,\omega)$ denoting the reflection response at $\setdD_0$
of the medium below $\setdD_0$.
At $\setdD_A$, we have $\u_B^\pm(\bx,\omega)=G^\pm(\bx,\bxr,\omega)$.
For state $A$ we consider a focal point at $\bxa=(x_{1,A},x_{3,A})$ at $\setdD_A$. The medium in state $A$ is a truncated medium, 
which is identical to the actual medium between $\setdD_0$ and $\setdD_A$, and homogeneous below $\setdD_A$. 
At $\setdD_0$ a downgoing focusing function $\u_A^+(\bx,\omega)=f_1^+(\bx,\bxa,\omega)$, with $\bx=(x_1,x_{3,0})$, is incident to the truncated medium.
This function focuses  at $\bxa$, hence, at $\setdD_A$ we have  $\u_A^+(\bx,\omega)=f_1^+(\bx,\bxa,\omega)=\delta(\bxh-\bxha)$.
The response to this focusing function at $\setdD_0$ is $\u_A^-(\bx,\omega)=f_1^-(\bx,\bxa,\omega)$.
Because the truncated medium is homogeneous below  $\setdD_A$,  we have  $\u_A^-(\bx,\omega)=0$ at $\setdD_A$.
The quantities for both states are summarised in Table 2.
 
 \begin{center}
{\small
{\noindent \it Table 2: Quantities to derive equations (\ref{eq145}) and (\ref{eq146}).}
\begin{tabular}{||l|c|c|c|c||}
\hline\hline
& $\u_A^+(\bx,\omega)$ & $\u_A^-(\bx,\omega)$ & $\u_B^+(\bx,\omega)$ & $\u_B^-(\bx,\omega)$  \\
\hline
$\bx=(x_1,x_{3,0})$ at $\setdD_0$
& $f_1^+(\bx,\bxa,\omega)$ &$f_1^-(\bx,\bxa,\omega)$ & $\delta(\bxh-\bxhr) $ & $R(\bx,\bxr,\omega)$\\
\hline
$\bx=(x_1,x_{3,A})$ at $\setdD_A$
& $\delta(\bxh-\bxha)$ &$0$ & $G^+(\bx,\bxr,\omega)$ & $G^-(\bx,\bxr,\omega)$ \\
\hline\hline
\end{tabular}
}
\end{center}
\mbox{}\\

Note that the downgoing focusing function $f_1^+(\bx,\bxa,\omega)$, for  $\bx$ at $\setdD_0$, is the inverse of the transmission response 
$T(\bxa,\bx,\omega)$ of the truncated medium \cite{Wapenaar2014JASA}, hence
\begin{eqnarray}\label{eqTinv}
f_1^+(\bx,\bxa,\omega)=T^{\rm inv}(\bxa,\bx,\omega),
\end{eqnarray}
for $\bx$ at $\setdD_0$.
To avoid instabilities in the evanescent field, the focusing function is in practice spatially band-limited.

Substituting the quantities of Table 2   into equations (\ref{eq142}) and (\ref{eq143}) gives 
\begin{eqnarray}
&&G^-(\bxa,\bxr,\omega)+f_1^{\am}(\bxr,\bxa,\omega)=
\int_{\setdD_0}R(\bx,\bxr,\omega)f_1^{\ap}(\bx,\bxa,\omega)\dx\label{eq145}
\end{eqnarray}
and
\begin{eqnarray}
&&G^+(\bxa,\bxr,\omega)-\{f_1^+(\bxr,\bxa,\omega)\}^*=
-\int_{\setdD_0}R(\bx,\bxr,\omega)\{f_1^-(\bx,\bxa,\omega)\}^*\dx,\label{eq146}
\end{eqnarray}
respectively. These are two representations for the upgoing and downgoing parts of the Green's function between $\bxr$ at the acquisition surface and $\bxa$ inside the non-reciprocal medium.
They are expressed in  terms of the reflection response  $R(\bx,\bxr,\omega)$ and a number of focusing functions.
Unlike similar representations for reciprocal media \cite{Slob2014GEO, Wapenaar2014JASA}, the focusing functions in equation (\ref{eq145}) are defined in the  \rev{complementary version} 
of the truncated medium.
Therefore we cannot use the standard approach to retrieve the focusing functions and Green's functions from the reflection response  $R(\bx,\bxr,\omega)$.
We obtain a second set of representations by replacing all quantities in equations (\ref{eq145}) and (\ref{eq146}) by the corresponding quantities in the  \rev{complementary} medium. 
For the focusing functions in equation (\ref{eq145}) this implies they are replaced by their counterparts in the truncated actual medium.  
We thus obtain
\begin{eqnarray}
&&G^{\am}(\bxa,\bxr,\omega)+f_1^-(\bxr,\bxa,\omega)=
\int_{\setdD_0}R^{\a}(\bx,\bxr,\omega)f_1^+(\bx,\bxa,\omega)\dx\label{eq145b}
\end{eqnarray}
and
\begin{eqnarray}
&&G^{\ap}(\bxa,\bxr,\omega)-\{f_1^{\ap}(\bxr,\bxa,\omega)\}^*=
-\int_{\setdD_0}R^{\a}(\bx,\bxr,\omega)\{f_1^{\am}(\bx,\bxa,\omega)\}^*\dx,\label{eq146b}
\end{eqnarray}
respectively. Because in practical situations we do not have access to the reflection response $R^{\a}(\bx,\bxr,\omega)$ 
 in the  \rev{complementary} medium, we derive a relation analogous to equation (\ref{eq11})
for this reflection response. To this end, consider the quantities in Table 3, 
with  $\bx_S$ and $\bx_R$ just above $\setdD_0$, and with $\setdD_M$ denoting a boundary below all inhomogeneities, so that there are no upgoing waves at $\setdD_M$.
Substituting the quantities of Table 3   into equation (\ref{eq142})  (with $\setdD_A$ replaced by $\setdD_M$) gives 
\begin{equation}\label{eq27qq}
R^{\a}(\bx_R,\bxr,\omega)=R(\bxr,\bx_R,\omega).
\end{equation}
Equations  (\ref{eq145}) $-$ (\ref{eq146b}), with $R^{\a}(\bx,\bxr,\omega)$ replaced by $R(\bxr,\bx,\omega)$, form the basis for the Marchenko method, discussed in the next section.\\

\begin{center}
{\small
{\noindent \it Table 3: Quantities to derive equation (\ref{eq27qq}).}
\begin{tabular}{||l|c|c|c|c||}
\hline\hline
& $\u_A^{\ap}(\bx,\omega)$ & $\u_A^{\am}(\bx,\omega)$ & $\u_B^+(\bx,\omega)$ & $\u_B^-(\bx,\omega)$  \\
\hline
$\bx=(x_1,x_{3,0})$ at $\setdD_0$
& $\delta(\bxh-\bxhr) $ & $R^{\a}(\bx,\bx_S,\omega)$  & $\delta(\bxh-x_{1,R}) $ & $R(\bx,\bx_R,\omega)$\\
\hline
$\bx=(x_1,x_{3,M})$ at $\setdD_M$
& $G^{\ap}(\bx,\bxr,\omega)$ &$0$ & $G^+(\bx,\bx_R,\omega)$ & 0 \\
\hline\hline
\end{tabular}
}
\end{center}

\section{The Marchenko method for non-reciprocal media}

The standard multidimensional Marchenko method for  reciprocal media \cite{Slob2014GEO, Wapenaar2014JASA} 
uses the representations of  equations (\ref{eq145}) and (\ref{eq146}), but without the superscript $\a$,
 to retrieve the focusing functions from the reflection response. Here we discuss how to  modify this method for non-reciprocal media. 
We  separate the representations of equations (\ref{eq145}) $-$ (\ref{eq146b}) into two sets, each set containing focusing functions in one and the same truncated medium. These sets are
equations (\ref{eq146}) and (\ref{eq145b}), with the focusing functions in the truncated actual medium, and equations (\ref{eq145}) and (\ref{eq146b}), 
with the focusing functions in the truncated  \rev{complementary} medium. We start with the set of equations (\ref{eq146}) and (\ref{eq145b}), which read in the time domain 
 (using equation (\ref{eq27qq}))
\begin{eqnarray}
&&G^+(\bxa,\bxr,t)-f_1^+(\bxr,\bxa,-t)=
-\int_{\setdD_0}\dx\int_{-\infty}^tR(\bx,\bxr,t-t')f_1^-(\bx,\bxa,-t')\dt\label{eq146t}
\end{eqnarray}
and
\begin{eqnarray}
&&G^{\am}(\bxa,\bxr,t)+f_1^-(\bxr,\bxa,t)=
\int_{\setdD_0}\dx\int_{-\infty}^t R(\bxr,\bx,t-t')f_1^+(\bx,\bxa,t')\dt,\label{eq145t}
\end{eqnarray}
respectively.  We introduce time windows to remove the Green's functions from these representations. 
Similar as in the reciprocal situation, 
we assume that the Green's function and the time-reversed focusing function on the left-hand side of equation (\ref{eq146t}) are separated in time, 
except for the direct arrivals \cite{Wapenaar2014JASA}. 
This is a reasonable assumption for media with smooth lateral variations, and for limited horizontal source-receiver distances.
Let $t_{\rm d}(\bxa,\bxr)$ denote the travel time of the direct arrival of $G^+(\bxa,\bxr,t)$. We define a time window 
$w(\bxa,\bxr,t)=u(t_{\rm d}(\bxa,\bxr)-t_\epsilon-t)$, where $u(t)$ is the Heaviside function and $t_\epsilon$ a small positive time constant. 
Under the above-mentioned assumption,  we have $w(\bxa,\bxr,t)G^+(\bxa,\bxr,t)=0$. For the focusing function on the left-hand side of
equation (\ref{eq146t}) we write \cite{Wapenaar2014JASA}
\begin{eqnarray}
f_1^+(\bxr,\bxa,t)&=&T^{\rm inv}(\bxa,\bxr,t)\nonumber\\
&=&T_{\rm d}^{\rm inv}(\bxa,\bxr,t)+M^+(\bxr,\bxa,t),\label{eqf1p}
\end{eqnarray}
where $T_{\rm d}^{\rm inv}(\bxa,\bxr,t)$ is the inverse of the direct arrival of the transmission response of the truncated medium and $M^+(\bxr,\bxa,t)$ the scattering coda.
The travel time of $T_{\rm d}^{\rm inv}(\bxa,\bxr,t)$ is  $-t_{\rm d}(\bxa,\bxr)$ and the scattering coda obeys $M^+(\bxr,\bxa,t)=0$ for $t\le-t_{\rm d}(\bxa,\bxr)$.
Hence, $w(\bxa,\bxr,t)f_1^+(\bxr,\bxa,-t)=M^+(\bxr,\bxa,-t)$.
Applying the time window $w(\bxa,\bxr,t)$ to both sides of equation (\ref{eq146t}) thus yields  
\begin{eqnarray}
&&M^+(\bxr,\bxa,-t)=
w(\bxa,\bxr,t)\int_{\setdD_0}\dx\int_{-\infty}^tR(\bx,\bxr,t-t')f_1^-(\bx,\bxa,-t')\dt.\label{eq146tw}
\end{eqnarray}

 \rev{Under the same conditions as those mentioned for equation (\ref{eq146t}),} 
we assume  that the Green's function and the focusing function in the left-hand side of equation (\ref{eq145t}) are separated in time (without overlap).
Unlike for  reciprocal media, we need a different time window to suppress the Green's function, because the latter is defined in the  \rev{complementary} medium.
To this end we define a time window  $w^{\a}(\bxa,\bxr,t)=u(t_{\rm d}^{\a}(\bxa,\bxr)-t_\epsilon-t)$, 
where  $t_{\rm d}^{\a}(\bxa,\bxr)$ denotes the travel time of the direct arrival in the  \rev{complementary} medium.
Applying this window to both sides of equation (\ref{eq145t}) yields
\begin{eqnarray}
&&f_1^-(\bxr,\bxa,t)=
w^{\a}(\bxa,\bxr,t)\int_{\setdD_0}\dx\int_{-\infty}^t R(\bxr,\bx,t-t')f_1^+(\bx,\bxa,t')\dt.\label{eq145tw}
\end{eqnarray}
Equations (\ref{eq146tw}) and (\ref{eq145tw}), with $f_1^+$ given by equation (\ref{eqf1p}), 
form a set of two equations for the two unknown functions $M^+(\bx,\bxa,t)$ and $f_1^-(\bx,\bxa,t)$ (with $\bx$ at $\setdD_0$). These functions can be resolved from
equations (\ref{eq146tw}) and (\ref{eq145tw}), assuming $R(\bx,\bxr,t)$, $R(\bxr,\bx,t)$,  $t_{\rm d}(\bxa,\bxr)$, $t_{\rm d}^{\a}(\bxa,\bxr)$ and $T_{\rm d}^{\rm inv}(\bxa,\bxr,t)$ 
 are known for all $\bx$ and $\bxr$ at $\setdD_0$. The reflection responses $R(\bx,\bxr,t)$ and $R(\bxr,\bx,t)$ are  obtained from measurements at the upper boundary $\setdD_0$ of the medium. 
This involves deconvolution for the source function, decomposition and, when the upper boundary is a reflecting boundary, 
elimination of the surface-related multiple reflections \cite{Verschuur92GEO}. 
 \rev{Because the deconvolution is limited by the bandwidth of the source function, the time constant $t_\epsilon$ in the window function is taken equal to half the duration of the source function.
This implies that the method will not account for short period multiples in layers with a thickness smaller than the wavelength \cite{Slob2014GEO}.
}
The travel times $t_{\rm d}(\bxa,\bxr)$ and $t_{\rm d}^{\a}(\bxa,\bxr)$, and the
 inverse of the direct arrival of the transmission response, $T_{\rm d}^{\rm inv}(\bxa,\bxr,t)$, can be derived from a background model of the medium and its  \rev{complementary version
 (once the background model is known, its complementary version follows immediately).}
 A smooth background model is sufficient to derive these quantities, hence, no information about the scattering interfaces inside the medium is required.
 The iterative Marchenko scheme to solve  for  $M^+(\bx,\bxa,t)$ and $f_1^-(\bx,\bxa,t)$ reads
 \begin{eqnarray}
f_{1,k}^-(\bxr,\bxa,t)&=&
w^{\a}(\bxa,\bxr,t)\int_{\setdD_0}\dx\int_{-\infty}^t R(\bxr,\bx,t-t')f_{1,k}^+(\bx,\bxa,t')\dt,\label{eq145twit}\\
M_{k+1}^+(\bxr,\bxa,-t)&=&
w(\bxa,\bxr,t)\int_{\setdD_0}\dx\int_{-\infty}^tR(\bx,\bxr,t-t')f_{1,k}^-(\bx,\bxa,-t')\dt,\label{eq146twit}
\end{eqnarray}
with
\begin{eqnarray}
f_{1,k}^+(\bx,\bxa,t)=T_{\rm d}^{\rm inv}(\bxa,\bx,t)+M_k^+(\bx,\bxa,t),\label{eqf1pit}
\end{eqnarray}
starting with $M_0^+(\bx,\bxa,t)=0$. Once $M^+(\bx,\bxa,t)$ and $f_1^-(\bx,\bxa,t)$  are found, $f_1^+(\bx,\bxa,t)$ is obtained from equation (\ref{eqf1p}) and, subsequently,
the Green's functions $G^+(\bxa,\bxr,t)$ and $G^{\am}(\bxa,\bxr,t)$ are obtained from equations (\ref{eq146t}) and (\ref{eq145t}).
Note that only $G^+(\bxa,\bxr,t)$ is defined in the actual medium. 
To obtain $G^-(\bxa,\bxr,t)$ in the actual medium we consider the set of equations
(\ref{eq145}) and (\ref{eq146b}), which read in the time domain (using equation (\ref{eq27qq}))
\begin{eqnarray}
&&G^-(\bxa,\bxr,t)+f_1^{\am}(\bxr,\bxa,t)=\int_{\setdD_0}\dx \int_{-\infty}^t R(\bx,\bxr,t-t')f_1^{\ap}(\bx,\bxa,t')\dt\label{eq145tt}
\end{eqnarray}
and
\begin{equation}
G^{\ap}(\bxa,\bxr,t)-f_1^{\ap}(\bxr,\bxa,-t)=-\int_{\setdD_0}\dx\int_{-\infty}^t R(\bxr,\bx,t-t')f_1^{\am}(\bx,\bxa,-t')\dt,\label{eq146bt}
\end{equation}
respectively.
The same reasoning as above leads to the following iterative Marchenko scheme for the focusing functions in the truncated  \rev{complementary} medium
\begin{eqnarray}
f_{1,k}^{\am}(\bxr,\bxa,t)&=&w(\bxa,\bxr,t)
\times\int_{\setdD_0}\dx \int_{-\infty}^t R(\bx,\bxr,t-t')f_{1,k}^{\ap}(\bx,\bxa,t')\dt\label{eq145ttaait}\\
M_{k+1}^{\ap}(\bxr,\bxa,-t)&=&w^{\a}(\bxa,\bxr,t)\int_{\setdD_0}\dx\int_{-\infty}^t R(\bxr,\bx,t-t')f_{1,k}^{\am}(\bx,\bxa,-t')\dt,\label{eq146btaait}
\end{eqnarray}
with
\begin{eqnarray}
f_{1,k}^{\ap}(\bx,\bxa,t)=T_{\rm d}^{{\rm inv}\a}(\bxa,\bx,t)+M_k^{\ap}(\bx,\bxa,t),\label{eqf1paait}
\end{eqnarray}
starting with $M_0^{\ap}(\bx,\bxa,t)=0$. Here $T_{\rm d}^{{\rm inv}\a}(\bxa,\bx,t)$ can be derived from the  \rev{complementary} background model.
Once the focusing functions $f_1^{\ap}(\bx,\bxa,t)$ and $f_1^{\am}(\bx,\bxa,t)$  are found, 
the Green's functions  $G^-(\bxa,\bxr,t)$ and $G^{\ap}(\bxa,\bxr,t)$ are obtained from  equations
(\ref{eq145tt}) and (\ref{eq146bt}).

\begin{center}
{\small
{\noindent \it Table 4: Quantities to derive equation (\ref{eqMDD}).}
\begin{tabular}{||l|c|c|c|c||}
\hline\hline
& $\u_A^{\ap}(\bx,\omega)$ & $\u_A^{\am}(\bx,\omega)$ & $\u_B^+(\bx,\omega)$ & $\u_B^-(\bx,\omega)$  \\
\hline
$\bx=(x_1,x_{3,A})$ at $\setdD_A$
& $\delta(\bxh-\bxha) $ & $R^{\a}(\bx,\bx_A,\omega)$  & $G^+(\bx,\bx_S,\omega)$ & $G^-(\bx,\bx_S,\omega)$\\
\hline
$\bx=(x_1,x_{3,M})$ at $\setdD_M$
& $G^{\ap}(\bx,\bx_A,\omega)$ &$0$ & $G^+(\bx,\bx_S,\omega)$ & 0 \\
\hline\hline
\end{tabular}
}
\end{center}
\mbox{}\\

We conclude this section by showing how $G^+(\bxa,\bxr,t)$ and $G^-(\bxa,\bxr,t)$ can be used to image the interior of the non-reciprocal medium.
First we derive a mutual relation between these Green's functions. To this end, consider the quantities in Table 4. Here $R^{\a}(\bx,\bx_A,\omega)$ in state $A$ is the reflection response
at $\setdD_A$ 
of the  \rev{complementary} medium below $\setdD_A$, with $\bx_A$ defined just above $\setdD_A$ and the medium in state $A$ being homogeneous above $\setdD_A$.
Substituting the quantities of Table 4   into equation (\ref{eq142})  (with $\setdD_0$ and $\setdD_A$ replaced by $\setdD_A$ and $\setdD_M$, respectively)  \rev{and using equation  (\ref{eq27qq}),} gives
\begin{eqnarray}
&&G^-(\bxa,\bxr,\omega)=
\int_{\setdD_A}R(\bxa,\bx,\omega)G^+(\bx,\bxr,\omega)\dx,\label{eqMDD}
\end{eqnarray}
or, applying an inverse Fourier transformation to the time domain,
\begin{eqnarray}
&&G^-(\bxa,\bxr,t)=
\int_{\setdD_A}\dx \int_{-\infty}^tR(\bxa,\bx,t-t')G^+(\bx,\bxr,t')\dt.\label{eqMDDt}
\end{eqnarray}
Given the Green's functions $G^+(\bx,\bxr,t)$ and $G^-(\bxa,\bxr,t)$ for all $\bxa$ and $\bx$ at $\setdD_A$ for a range of source positions $\bxr$ at $\setdD_0$, the reflection response
$R(\bxa,\bx,t)$ for all $\bxa$ and $\bx$ at $\setdD_A$ can be resolved by multidimensional deconvolution \cite{Wapenaar2000SEG, Amundsen2001GEO, Holvik2005GEO, Wapenaar2010JASA, Neut2011GEO, Ravasi2015GJI}. An image can be obtained by selecting $R(\bxa,\bxa,t=0)$  and repeating the process for any $\bxa$ in the region of interest.

 \rev{ 
We discuss an alternative imaging approach for the special case of a laterally invariant medium.
To this end we first rewrite equation (\ref{eqMDD}) as a spatial convolution, taking $x_{1,S}=0$, hence
\begin{eqnarray}
&&G^-(x_{1,A},x_{3,A},x_{3,S},\omega)=
\int_{-\infty}^\infty R(x_{1,A}-x_1,x_{3,A},\omega)G^+(x_1,x_{3,A},x_{3,S},\omega)\dxx.\label{eqMDDhh}
\end{eqnarray}
We define the spatial Fourier transform of a function $\f(x_1,x_3,\omega)$ as
\begin{equation}\label{eqFTP}
\tilde \f(\p,x_3,\omega)=\int_{-\infty}^\infty \f(x_1,x_3,\omega)\exp(-\i\omega \p x_1)\dxx,
\end{equation}
with $\p$ being the horizontal slowness. In the $(\p,x_3,\omega)$-domain, equation (\ref{eqMDDhh}) becomes 
\begin{equation}
\tilde G^-(s_1,x_{3,A},x_{3,S},\omega)=\tilde R(s_1,x_{3,A},\omega)\tilde G^+(s_1,x_{3,A},x_{3,S},\omega),
\end{equation}
or, applying an inverse Fourier transformation to the time domain,
\begin{equation}\label{eq47}
G^-(s_1,x_{3,A},x_{3,S},\tau)=\int_{-\infty}^\tau R(s_1,x_{3,A},\tau-\tau')G^+(s_1,x_{3,A},x_{3,S},\tau'){\rm d}\tau'.
\end{equation}
Given the Green's functions $G^+ (s_1,x_{3,A},x_{3,S},\tau) $ and $G^-(s_1,x_{3,A},x_{3,S},\tau)$, the reflection response
$R(s_1,x_{3,A},\tau)$ for each horizontal slowness $s_1$ can be resolved by 1D  deconvolution. An image can be obtained by selecting $R(s_1,x_{3,A},\tau=0)$  and repeating the process 
for all $s_1$ and for any  $x_{3,A}$ in the region of interest.
}

\begin{figure}
\centerline{\epsfysize=11 cm \epsfbox{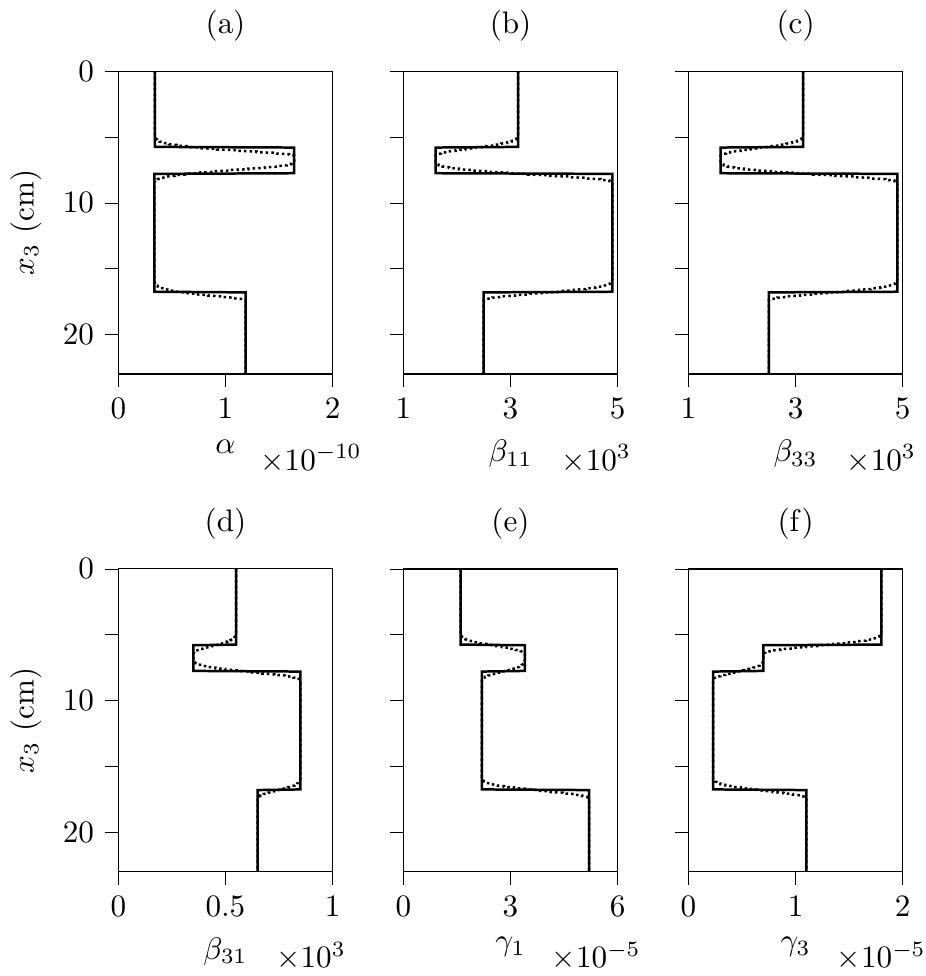}}
\caption{\footnotesize  \rev{Solid lines:} parameters $\alpha(x_3)$, $\beta_{11}(x_3)$, $\beta_{33}(x_3)$, $\beta_{31}(x_3)$, $\gamma_1(x_3)$ and $\gamma_3(x_3)$ of the layered medium.
 \rev{Dotted lines: smoothed medium parameters, used to model the initial estimate of the focusing functions}.}
\label{Fig11}
\end{figure}

 \begin{figure}
\centerline{\epsfysize=10 cm \epsfbox{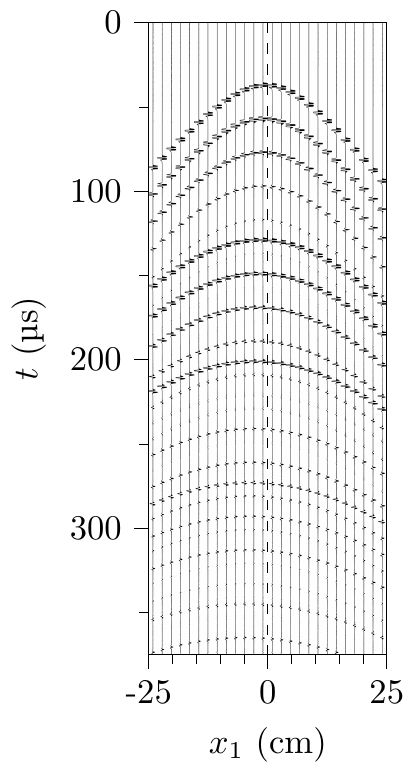}}
\caption{\footnotesize The modeled reflection response $R(\bx,\bxr,t)*S(t)$ at $\setdD_0$.  \rev{Note the asymmetry with respect to the dashed line due to the non-reciprocal medium parameters.}}
\label{Fig11b}
\end{figure}

\section{Numerical example}

\begin{figure}
\centerline{\epsfysize=14 cm \epsfbox{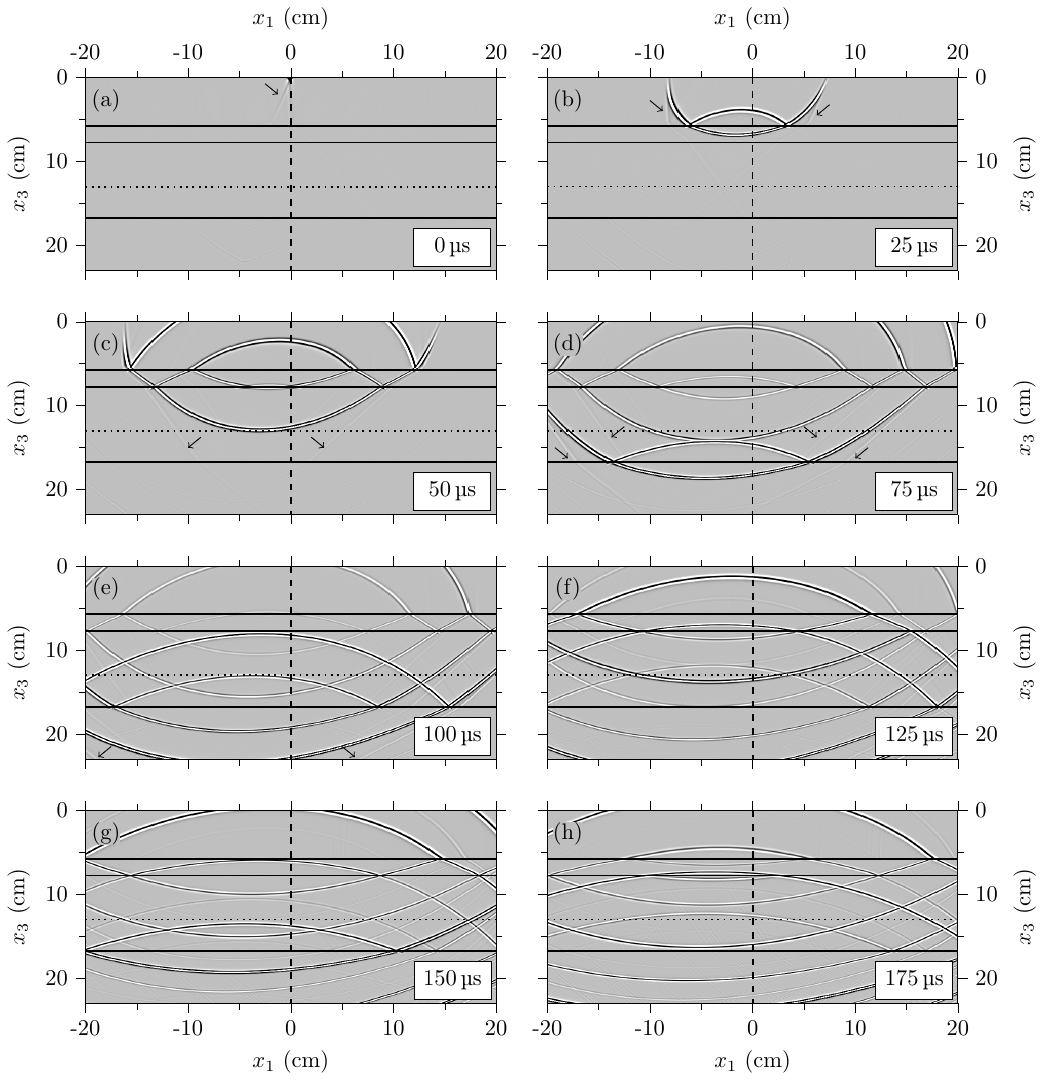}}
\caption{\footnotesize Snapshots of $\{G^+(\bxa,\bxr,t)+G^-(\bxa,\bxr,t)\}*S(t)$, retrieved via equations (\ref{eq146t}) and (\ref{eq145tt}), for $\bxr=(0,0)$ and variable $\bxa$.}
\label{Fig12}
\end{figure}

We illustrate the proposed methodology with a numerical example,  \rev{mimicking an ultrasound experiment}. For simplicity we consider a horizontally layered medium, 
consisting of three homogeneous layers and a homogeneous half-space below the deepest layer. The medium parameters of the layered medium,
$\alpha(x_3)$, $\beta_{rs}(x_3)$ and $\gamma_r(x_3)$ are shown in Figure \ref{Fig11}.
In many practical situations the parameters  $\beta_{31}(x_3)$ and $\gamma_3(x_3)$ will be zero, 
but we choose them to be non-zero to demonstrate the generality of the method. We define a source at $\bxr=(0,0)$ at the top of the first layer,
which emits a time-symmetric wavelet $S(t)$ with a central frequency of 600 kHz into the layered medium. We use
a wavenumber-frequency domain modelling method \cite{Kennett79GJRAS}, adjusted for non-reciprocal media, to model the response to this source.
The modelled reflection response, 
$R(\bx,\bxr,t)*S(t)$ at $\setdD_0$ (the asterisk denoting convolution), is shown in Figure \ref{Fig11b}. To emphasise the multiple scattering, a time-dependent
amplitude gain has been applied, using the function $\exp\{3t/375\mu s\}$.
Note that the apices of the reflection hyperbolae drift to the left with increasing time, which is a manifestation of the non-reciprocal medium parameters. 
Because the medium is laterally invariant, the response to any other source at the surface is just a laterally shifted version of the response in Figure \ref{Fig11b}.
We apply the Marchenko method, discussed in detail in the previous section, to derive the focusing functions $f_1^\pm(\bxr,\bxa,t)$ and $f_1^{\pm\a}(\bxr,\bxa,t)$
for fixed $\bxr=(0,0)$ and variable $\bxa$. As input we use the reflection response $R(\bx,\bxr,t)*S(t)$  \rev{of the actual medium}
and the direct arrivals $T_{\rm d}(\bxa,\bx,t)$ and $T_{\rm d}^{\a}(\bxa,\bx,t)$, modelled in  \rev{a smoothed version of the truncated} medium and its  \rev{complementary version
(the smoothed medium is indicated by the dotted lines in Figure \ref{Fig11}). 
For simplicity we approximate the inverse direct arrivals $T_{\rm d}^{\rm inv}(\bxa,\bx,t)$ and $T_{\rm d}^{{\rm inv}\a}(\bxa,\bx,t)$ 
in equations (\ref{eqf1pit}) and (\ref{eqf1paait}) by the time-reversals
$T_{\rm d}(\bxa,\bx,-t)$ and $T_{\rm d}^{\a}(\bxa,\bx,-t)$.} 
For $t_\epsilon$ in the time windows $w(\bxa,\bxr,t)$ and $w^{\a}(\bxa,\bxr,t)$ we choose half the duration of the symmetric  wavelet $S(t)$,
 i.e., $t_\epsilon=0.65 \mu$s, and the Heaviside functions are tapered. 
Because we consider a laterally invariant medium, the integrals in the right-hand sides of equations
(\ref{eq145twit}), (\ref{eq146twit}), (\ref{eq145ttaait}) and (\ref{eq146btaait}) are efficiently replaced by multiplications in the wavenumber-frequency domain.
In total we apply  \rev{20}  iterations of the Marchenko scheme to derive the focusing functions
 $f_1^\pm(\bxr,\bxa,t)*S(t)$ and the same number of iterations to derive  $f_1^{\pm\a}(\bxr,\bxa,t)*S(t)$.
These focusing functions are substituted into equations (\ref{eq146t}) and (\ref{eq145tt}) 
(of which the integrals are also evaluated via the wavenumber-frequency domain) to obtain the  \rev{wave fields} $G^+(\bxa,\bxr,t)*S(t)$ and $G^-(\bxa,\bxr,t)*S(t)$.
The superposition of these  \rev{wave fields}  is shown in grey-level display 
in Figure \ref{Fig12} in the form of snapshots (i.e., wave fields at frozen time), 
for fixed $\bxr=(0,0)$ and variable $\bxa$. 
 \rev{The amplitudes are clipped at 8$\%$ of the maximum amplitude.}
 This figure clearly shows the propagation of the wave field from the source through the layered non-reciprocal medium.
 The wavefronts are asymmetric as a result of the non-reciprocal medium parameters 
(for a reciprocal medium these snapshots would be symmetric with respect to the vertical dashed lines).
Multiple scattering between the layer interfaces is also clearly visible. The interfaces, indicated by the solid horizontal lines in each of the panels in  
Figure \ref{Fig12}, are only shown here to aid the interpretation of the retrieved Green's functions. 
However, no explicit information of these interfaces has been used to retrieve these Green's functions; all information about the scattering at the layer interfaces
comes directly from the reflection response $R(\bx,\bxr,t)*S(t)$. 
The snapshots also exhibit some weak spurious linear events  \rev{(indicated by the arrows in Figure \ref{Fig12})}, which are mainly caused by the negligence of 
evanescent waves and the absence of very large propagation angles in the reflection response.

\begin{figure}
\centerline{\epsfysize=9 cm  \epsfbox{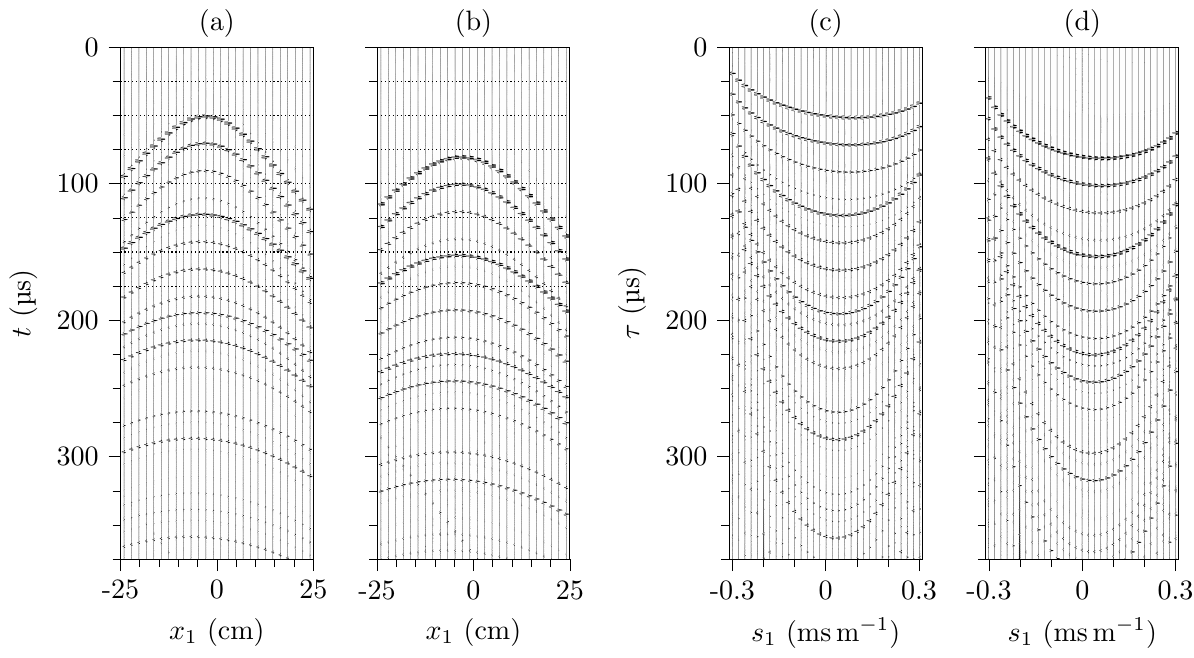}}
\caption{\footnotesize  \rev{Downgoing and upgoing wave fields at $x_{3,A}=13$ cm. (a) $G^+(x_1,x_{3,A},x_{3,S},t)*S(t)$, (b) $G^-(x_1,x_{3,A},x_{3,S},t)*S(t)$, 
(c) $G^+(s_1,x_{3,A},x_{3,S},\tau)*S(\tau)$, (d) $G^-(s_1,x_{3,A},x_{3,S},\tau)*S(\tau)$}.}
\label{Fig213}
\end{figure}

\begin{figure}
\centerline{\epsfysize=9 cm  \epsfbox{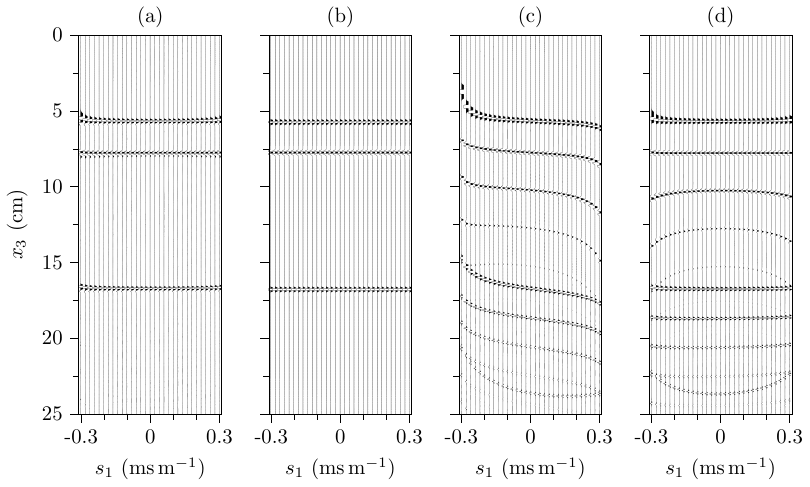}}
\caption{\footnotesize  \rev{Images in the $(s_1,x_3)$-domain of the layered medium of Figure \ref{Fig11}. 
(a) Marchenko imaging, accounting for non-reciprocity. (b) Reference reflectivity.
(c) Primary imaging, ignoring non-reciprocity. (d) Primary imaging, accounting for non-reciprocity. }}
\label{Fig214}
\end{figure}

Next, we image the interfaces of the layered medium,  \rev{following the approach for a laterally invariant medium described at the end of the previous section.
Figures \ref{Fig213}a,b show the downgoing and upgoing wave fields $G^+(x_1,x_{3,A},x_{3,S},t)*S(t)$ and $G^-(x_1,x_{3,A},x_{3,S},t)*S(t)$, respectively, for
$x_{3,A}=13$ cm (the depth of the horizontal dotted lines in Figure \ref{Fig12}). The horizontal dotted lines in Figures \ref{Fig213}a,b indicate the times of the snapshots
in Figure \ref{Fig12}. Figures \ref{Fig213}c,d show the downgoing and upgoing wave fields $G^+(s_1,x_{3,A},x_{3,S},\tau)*S(\tau)$ and $G^-(s_1,x_{3,A},x_{3,S},\tau)*S(\tau)$, respectively,
for a range of horizontal slownesses $s_1$. From these wave fields  we derive the reflection response $R(s_1,x_{3,A},\tau)$  by inverting
equation (\ref{eq47}) for each horizontal slowness $s_1$. The image at $x_{3,A}$ is obtained as $R(s_1,x_{3,A},\tau=0)$. We repeat this for all $x_{3,A}$ between 0 and 25 cm, in steps of 0.25 mm. The result is shown in Figure 
\ref{Fig214}a. This figure clearly shows images of the three interfaces in Figure \ref{Fig11}. For comparison, Figure \ref{Fig214}b shows, as a reference, the true reflectivity.
The relative amplitude errors of the imaged interfaces are between 0.5$\%$ and 2$\%$, except for slownesses $|s_1|>0.2$ ms/m, close to the evanescent field. 
Figure \ref{Fig214}c shows the result of standard primary imaging, ignoring non-reciprocity. The trace at $s_1=0$
contains images of the three interfaces at the correct depths, but it also contains false images caused by the internal multiples. Moreover, the traces
for $s_1\ne 0$ contain images at wrong depths only. Finally, Figure \ref{Fig214}d is the result of primary imaging, taking non-reciprocity into account (by applying one iteration 
with our method). The three interfaces are imaged at the correct depths for all horizontal slownesses, but the false images are not suppressed.} 

\section{Conclusions}

Marchenko imaging has recently been introduced as a novel approach to account for multiple scattering in multidimensional acoustic and electromagnetic imaging.
Given the recent interest in non-reciprocal materials, here we have extended the Marchenko approach for non-reciprocal media.
We have derived two iterative Marchenko schemes, one to retrieve focusing functions in
a truncated version of the actual medium and one to retrieve these functions in a truncated version of the  \rev{complementary} medium.
Both schemes use the reflection response of the actual medium as input, plus estimates of the direct arrivals of the transmission response of the truncated actual medium (for the 
first scheme) and of the truncated  \rev{complementary} medium (for the second scheme). We have derived Green's function representations, which express the downgoing and upgoing part of 
the Green's function inside the non-reciprocal medium, in terms of the reflection response at the surface of the actual medium and the focusing functions in the truncated 
actual and  \rev{complementary} medium. From these downgoing and upgoing Green's functions, a reflectivity image of the medium can be obtained.
We have illustrated the proposed approach at the hand of a numerical example for a horizontally layered non-reciprocal medium. 
This example shows an accurate  \rev{wave field}, propagating through the medium and scattering at its interfaces, retrieved from the reflection response at the surface. 
Moreover,  it shows an accurately obtained artefact-free reflectivity image of the non-reciprocal medium, which confirms that the proposed method properly
handles internal multiple scattering in a non-reciprocal medium.

\section*{Acknowledgements}
We thank  \rev{our colleague} Evert Slob for his advise about electromagnetic waves in non-reciprocal media 
 \rev{and reviewers Patrick Elison and Ivan Vasconcelos for their constructive comments, which helped to improve the paper.}
This work has received funding from the European Union's Horizon 2020 research and innovation programme: European Research Council (grant agreement 742703) and 
Marie Sk\l odowska-Curie (grant agreement 641943).

\appendix

\section{Wave equations for non-reciprocal media}\label{AppA}

We discuss wave equations for non-reciprocal media for (1) electromagnetic waves, (2) elastodynamic waves, and (3) acoustic waves. 
Next, (4) we derive a unified scalar wave equation for non-reciprocal media.

\subsection{Electromagnetic waves}

We start with the Maxwell equations for electromagnetic waves, 
\begin{eqnarray}
&&\partial_tD_i-\epsilon_{ijk}\partial_jH_k=-J_i^{\rm e},\label{eqa1}\\
&&\partial_tB_j+\epsilon_{jkl}\partial_kE_l=-J_j^{\rm m}.\label{eqa2}
\end{eqnarray}
Lower-case subscripts take the values 1, 2 and 3 and Einstein's summation convention applies to repeated subscripts. Exceptions are made for subscripts $r$, $s$ and $u$, which only take the values 1 and 3,
and for subscript $t$, which denotes time.
In equations (\ref{eqa1}) and (\ref{eqa2}), 
$E_l=E_l(\bx,t) $ is the electric field strength, 
$H_k=H_k(\bx,t)$ the magnetic field strength, 
$D_i=D_i(\bx,t)$ the electric flux density, 
$B_j=B_j(\bx,t)$ the magnetic flux density, 
$J_i^{\rm e}=J_i^{\rm e}(\bx,t)$ and $J_j^{\rm m}=J_j^{\rm m}(\bx,t)$ are source functions in terms of external electric and magnetic current densities and, finally, 
$\epsilon_{ijk}$ is the alternating tensor (or Levi-Civita tensor), with $\epsilon_{123}=\epsilon_{312}=\epsilon_{231}=1$,
$\epsilon_{213}=\epsilon_{321}=\epsilon_{132}=-1$, and all other components being zero. 
For metamaterials,  the field and source quantities in equations (\ref{eqa1}) and (\ref{eqa2}) are macroscopic quantities. 
These are sometimes denoted as $\langle H_k\rangle$ etc. \cite{Willis2011RS}, but for notational convenience we drop the brackets.
 \rev{In the low-frequency limit,} the effective constitutive relations for lossless metamaterials read \cite{Kong72IEEE, Kiehn91PRA, Willis2011RS}
\begin{eqnarray}
&&D_i=\varepsilon_{ij}E_j+\eta_{ij}B_j,\label{eqa3}\\
&&H_k=\theta_{kl}E_l+\nu_{kl}B_l,\label{eqa4}
\end{eqnarray}
where 
$\varepsilon_{ij}=\varepsilon_{ij}(\bx)$ is the permittivity,  
$\nu_{kl}=\nu_{kl}(\bx)$ the inverse permeability,
and $\eta_{ij}=\eta_{ij}(\bx)$ and $\theta_{kl}=\theta_{kl}(\bx)$ are coupling parameters.
 The inverse permeability is related to the permeability  $\mu_{jk}=\mu_{jk}(\bx)$ via
\begin{equation}\label{eqmunu}
\mu_{jk}\nu_{kl}=\delta_{jl},
\end{equation}
with  $\delta_{jl}$ the Kronecker delta function.
The medium parameters in equations (\ref{eqa3}) and (\ref{eqa4}) are effective parameters.
In general they are anisotropic, even when they are isotropic at micro scale.
For a non-reciprocal lossless metamaterial they are real-valued and obey the following symmetry relations \cite{Birss67PM, Kong72IEEE, Slob2009PIER}
\begin{equation}\label{eqepsnu}
\varepsilon_{ij}=\varepsilon_{ji}, \quad \nu_{kl}=\nu_{lk}, \quad \quad \mu_{jk}=\mu_{kj}, \quad \eta_{ij}=-\theta_{ji}. 
\end{equation}
We reorganise the constitutive relations into a set of explicit expressions for $D_i$ and $B_j$.
To this end we multiply both sides of equation (\ref{eqa4}) by $\mu_{jk}$. Using equation (\ref{eqmunu}) this gives
\begin{eqnarray}
&&B_j=-\mu_{jk}\theta_{kl}E_l+ \mu_{jk}H_k.\label{eqa5}
\end{eqnarray}
Substitution into equation (\ref{eqa3}) gives
\begin{eqnarray}
&&D_i=\bigl(\varepsilon_{il}-\eta_{ij}\mu_{jk}\theta_{kl}\bigr)E_l+\eta_{ij} \mu_{jk}H_k.\label{eqa6}
\end{eqnarray}
Equations (\ref{eqa6}) and (\ref{eqa5}) form a new set of effective constitutive relations \cite{Lindell95JEVA, Slob2012IEEE},
\begin{eqnarray}
&&D_i={\bareps}_{il}E_l+\xi_{ik}H_k,\label{eqa7}\\
&&B_j=\zeta_{jl}E_l+ \mu_{jk}H_k,\label{eqa8}
\end{eqnarray}
with
\begin{eqnarray}
&&{\bareps}_{il}=\varepsilon_{il}-\eta_{ij}\mu_{jk}\theta_{kl},\label{eqa9}\\
&&\xi_{ik}=\eta_{ij} \mu_{jk},\label{eqa10}\\
&&\zeta_{jl}=-\mu_{jk}\theta_{kl}.\label{eqa11}
\end{eqnarray}
On account of equation (\ref{eqepsnu}), these parameters obey the following symmetry relations \cite{Tellegen48PRR, Kong72IEEE}
\begin{equation}\label{eqepsnua}
{\bareps}_{il}={\bareps}_{li}, \quad \xi_{lj}=\zeta_{jl}. 
\end{equation}
Substitution of constitutive relations (\ref{eqa7}) and (\ref{eqa8}) into Maxwell equations (\ref{eqa1}) and (\ref{eqa2}), using $\xi_{lj}=\zeta_{jl}$, gives
\begin{eqnarray}
&&{\bareps}_{il}\partial_tE_l+\xi_{ik}\partial_tH_k-\epsilon_{ijk}\partial_jH_k=-J_i^{\rm e},\label{eqa12}\\
&&\xi_{lj}\partial_tE_l+\mu_{jk}\partial_tH_k+ \epsilon_{jkl}\partial_kE_l=-J_j^{\rm m}.\label{eqa13}
\end{eqnarray}
Next, we assume that the wave fields, sources and medium parameters are independent of the $x_2$-coordinate.
 Furthermore, we assume  
${\bareps}_{21}={\bareps}_{23}=0$, 
$\mu_{21}=\mu_{23}=0$, 
$\xi_{11}=\xi_{22}=\xi_{33}=\xi_{13}=\xi_{31}=0$.  
Then equation (\ref{eqa12}) for $i=1,2,3$ (using ${\bareps}_{13}={\bareps}_{31}$) and equation (\ref{eqa13}) for $j=1,2,3$ (using $\mu_{13}=\mu_{31}$) 
yield six equations, describing wave propagation in the $(x_1,x_3)$-plane.
These can be separated into two independent sets of equations, for transverse-electric (TE) waves (with wave field quantities  $E_2$, $H_1$ and $H_3$)
 and for transverse-magnetic (TM) waves (with wave field quantities  $H_2$, $E_1$ and $E_3$). 
For TE wave propagation in the $(x_1,x_3)$-plane we thus obtain
\begin{eqnarray}
&&{\bareps}_{22}\partial_tE_2 + \xi_{21}\partial_tH_1 + \xi_{23}\partial_tH_3 +\partial_1H_3 - \partial_3H_1=-J_2^{\rm e},\label{eqTE1}\\
&& \mu_{11}\partial_tH_1+  \mu_{31}\partial_tH_3 + \xi_{21}\partial_tE_2 - \partial_3E_2=-J_1^{\rm m},\label{eqTE2}\\
&& \mu_{31}\partial_tH_1 + \mu_{33}\partial_tH_3 + \xi_{23}\partial_tE_2 + \partial_1E_2=-J_3^{\rm m}\label{eqTE3}
\end{eqnarray}
and for TM wave propagation in the $(x_1,x_3)$-plane
\begin{eqnarray}
&&\mu_{22}\partial_tH_2 + \xi_{12}\partial_tE_1 + \xi_{32}\partial_tE_3 -\partial_1E_3 + \partial_3E_1=-J_2^{\rm m},\label{eqTM1}\\
&&{\bareps}_{11}\partial_tE_1+{\bareps}_{31}\partial_tE_3+\xi_{12}\partial_tH_2 + \partial_3H_2=-J_1^{\rm e},\label{eqTM2}\\
&&{\bareps}_{31}\partial_tE_1+{\bareps}_{33}\partial_tE_3+\xi_{32}\partial_tH_2 - \partial_1H_2=-J_3^{\rm e}.\label{eqTM3}
\end{eqnarray}

\subsection{Elastodynamic waves}

We start with the   \rev{equilibrium of momentum \cite{Nassar2017JMPS}} and the  \rev{deformation equation \cite{Hoop95Book}}
\begin{eqnarray}
&&\partial_t\m_i-\partial_j\tau_{ij}=\F_i,\label{eqb1}\\
&&\partial_t e_{kl}-\half(\partial_kv_l+\partial_lv_k)=-h_{kl}.\label{eqb2}
\end{eqnarray}
Here 
$\m_i=\m_i(\bx,t)$ is the momentum density, 
$\tau_{ij}=\tau_{ij}(\bx,t)$ the stress tensor, 
$e_{kl}=e_{kl}(\bx,t)$ the strain tensor,
$v_k=v_k(\bx,t)$ the particle velocity and 
$\F_i=\F_i(\bx,t)$ and $h_{kl}=h_{kl}(\bx,t)$ are source functions in terms of external force and deformation-rate density.
For metamaterials, the field and source quantities in equations (\ref{eqb1}) and (\ref{eqb2}) are macroscopic quantities.
These are sometimes denoted as $\langle \tau_{ij} \rangle$ etc. \cite{Willis2012CRM}, but for notational convenience we drop the brackets.
They obey the following symmetry relations
\begin{equation}\label{eqekllk}
\tau_{ij}=\tau_{ji},\quad e_{kl}=e_{lk},\quad h_{kl}=h_{lk}.
\end{equation}
 \rev{In the low-frequency limit,} the effective constitutive relations for metamaterials read \cite{Willis2012CRM, Norris2012RS, Nassar2017JMPS}
\begin{eqnarray}
&&\m_i=\rho_{ik} v_k+\etaa_{ikl}e_{kl},\label{eqb3}\\
&&\tau_{mn}=\thetaa_{mnp}v_p+c_{mnpq}e_{pq},\label{eqb4}
\end{eqnarray}
where $\rho_{ik}=\rho_{ik}(\bx)$ is the mass density tensor, 
$c_{mnpq}=c_{mnpq}(\bx)$ the stiffness tensor
and $\thetaa_{mnp}=\thetaa_{mnp}(\bx)$ and  $\etaa_{ikl}=\etaa_{ikl}(\bx)$  are coupling parameters. 
The stiffness tensor is related to the compliance tensor $s_{klmn}=s_{klmn}(\bx)$ via
\begin{equation}\label{eqscsym}
s_{klmn}c_{mnpq}=\half(\delta_{kp}\delta_{lq}+\delta_{kq}\delta_{lp}).
\end{equation}
The medium parameters in equations (\ref{eqb3}) and (\ref{eqb4}) are effective parameters. In general they are anisotropic, even when they are isotropic at micro scale.
An example of a non-reciprocal metamaterial is a phononic crystal of which the stiffness and mass density are modulated in a wave-like fashion \cite{Nassar2017JMPS}.
For this situation, equations (\ref{eqb3}) and (\ref{eqb4}) are defined in a coordinate system that moves along with the modulating wave, so that the effective medium parameters in
this coordinate system are time-independent.
For a non-reciprocal lossless metamaterial the medium parameters are real-valued and obey the following symmetry relations \cite{Nassar2017JMPS}
\begin{eqnarray}
&&\rho_{ik}=\rho_{ki},\label{eqA29}\\
&&c_{mnpq}=c_{nmpq}=c_{mnqp}=c_{pqmn},\\
&&s_{klmn}=s_{lkmn}=s_{klnm}=s_{mnkl},\label{eqA31}\\
&&\thetaa_{mnp}=\thetaa_{nmp},\\
&&\etaa_{ikl}=\etaa_{ilk}, \\
&&\etaa_{ikl}=-\thetaa_{kli}.\label{eqA34}
\end{eqnarray}
We reorganise the constitutive relations into a set of explicit expressions for $\m_i$ and $e_{kl}$.
To this end we multiply both sides of equation (\ref{eqb4}) by $s_{klmn}$. Using equations (\ref{eqscsym}) and $e_{kl}=e_{lk}$ this gives
\begin{eqnarray}
&&e_{kl}=-s_{klmn}\thetaa_{mnp}v_p+s_{klmn}\tau_{mn}.\label{eqb5}
\end{eqnarray}
Substitution into equation (\ref{eqb3}) gives
\begin{eqnarray}
\m_i=(\rho_{ip}-\etaa_{ikl}s_{klmn}\thetaa_{mnp})v_p+\etaa_{ikl}s_{klmn}\tau_{mn}.\label{eqb6}
\end{eqnarray}
Equations (\ref{eqb6}) and (\ref{eqb5}) form a new set of effective constitutive relations,
\begin{eqnarray}
&&\m_i={\barrho}_{ip}v_p-\xi_{imn}\tau_{mn},\label{eqb7}\\
&&e_{kl}=-\zeta_{klp}v_p+s_{klmn}\tau_{mn},\label{eqb8}
\end{eqnarray}
with
\begin{eqnarray}
&&{\barrho}_{ip}=\rho_{ip}-\etaa_{ikl}s_{klmn}\thetaa_{mnp},\label{eqb9}\\
&&\xi_{imn}=-\etaa_{ikl}s_{klmn},\label{eqb10}\\
&&\zeta_{klp}=s_{klmn}\thetaa_{mnp}.\label{eqb11}
\end{eqnarray}
For convenience we use the same symbols ($\xi$ and $\zeta$) for the coupling parameters as in the electromagnetic constitutive relations, but of course these are different quantities with different physical dimensions.
On account of equations (\ref{eqA29}), (\ref{eqA31}) and (\ref{eqA34}) these parameters obey the following  symmetry relations
\begin{equation}
{\barrho}_{ip}={\barrho}_{pi},\quad \zeta_{klp}=\zeta_{lkp},\quad \xi_{imn}=\xi_{inm}, \quad \xi_{pkl}=\zeta_{klp}.
\end{equation}
Substitution of constitutive relations (\ref{eqb7}) and (\ref{eqb8}) into equations (\ref{eqb1}) and (\ref{eqb2}), using $ \xi_{pkl}=\zeta_{klp}$, gives 
\begin{eqnarray}
&&{\barrho}_{ip}\partial_tv_p-\xi_{imn}\partial_t\tau_{mn}-\partial_j\tau_{ij}=\F_i,\label{eqb12}\\
&&-\xi_{pkl}\partial_tv_p+s_{klmn}\partial_t\tau_{mn}-\half(\partial_kv_l+\partial_lv_k)=-h_{kl}.\label{eqb13}
\end{eqnarray}
Next, we assume that the wave fields, sources and medium parameters are independent of the $x_2$-coordinate.
Furthermore, we assume 
${\barrho}_{21}={\barrho}_{23}=0$,
$s_{1211}=s_{1222}=s_{1233}=s_{1213}=s_{3211}=s_{3222}=s_{3233}=s_{3213}=0$ and
$\xi_{112}=\xi_{132}=
\xi_{211}=\xi_{222}=\xi_{233}=\xi_{213}=\xi_{312}=\xi_{332}=0$.
Then equation (\ref{eqb12}) for $i=2$ (using $\xi_{2mn}=\xi_{2nm}$ and $\tau_{mn}=\tau_{nm}$) 
and equation (\ref{eqb13}) for $k=1,3$ (setting $l=2$ in both cases and using equation (\ref{eqA31})) yield three equations, describing the propagation
 of horizontally-polarised shear (SH) waves
 (with wave field quantities  $v_2$, $\tau_{21}$ and $\tau_{23}$) in the $(x_1,x_3)$-plane:
 \begin{eqnarray}
&&{\barrho}_{22}\partial_tv_2 - 2\xi_{221}\partial_t\tau_{21} - 2\xi_{223}\partial_t\tau_{23} -\partial_1\tau_{21} - \partial_3\tau_{23}=\F_2,\label{eqSH1}\\
&& -4s_{1221}\partial_t\tau_{21}-4s_{1223}\partial_t\tau_{23} +2\xi_{221}\partial_tv_2 + \partial_1v_2=2h_{21},\label{eqSH2}\\
&& -4s_{1223}\partial_t\tau_{21}-4s_{3223}\partial_t\tau_{23} +2\xi_{223}\partial_tv_2 + \partial_3v_2=2h_{23}.\label{eqSH3}
\end{eqnarray}

\subsection{Acoustic waves}

We derive the equations for acoustic waves from those for elastodynamic waves. 
To this end we make the following substitutions
\begin{eqnarray}
\tau_{ij}&=&-\delta_{ij}\sigmaa,\\
e_{kl}&=&\third\delta_{kl}\Theta,\\
h_{kl}&=&\third\delta_{kl}q,\\
c_{mnpq}&=&\delta_{mn}\delta_{pq}\K.
\end{eqnarray}
Here 
$\sigmaa=\sigmaa(\bx,t)$ is the acoustic pressure, 
$\Theta=\Theta(\bx,t)$ the cubic dilatation, 
$q=q(\bx,t)$ a source function in terms of volume injection-rate density and
$\K=\K(\bx)$ the effective bulk modulus of the medium.
With these substitutions, equations (\ref{eqb1}) and (\ref{eqb2}) become
\begin{eqnarray}
&&\partial_t\m_i+\partial_i\sigmaa=\F_i,\label{eqbb1}\\
&&\third\delta_{kl}\partial_t \Theta-\half(\partial_kv_l+\partial_lv_k)=-\third\delta_{kl}q.\label{eqbbb2}
\end{eqnarray}
Multiplying both sides of the latter equation by $\delta_{kl}$ we obtain
\begin{eqnarray}
&&\partial_t \Theta-\partial_kv_k=-q.\label{eqbb2}
\end{eqnarray}
Similarly, the constitutive relations (\ref{eqb3}) and (\ref{eqb4}) become
\begin{eqnarray}
&&\m_i=\rho_{ik} v_k+\third\etaa_{ill}\Theta,\label{eqbb3}\\
&&-\delta_{mn}\sigmaa=\thetaa_{mnp}v_p+\third\delta_{mn}\delta_{pq}\K\delta_{pq}\Theta.\label{eqbbb4}
\end{eqnarray}
Multiplying both sides of the latter equation by $\third\delta_{mn}$ we obtain
\begin{eqnarray}
&&-\sigmaa=\third\thetaa_{mmp}v_p+\K\Theta.\label{eqbbc4}
\end{eqnarray}
On account of equations (\ref{eqA29}) and (\ref{eqA34}), the effective medium parameters in constitutive relations (\ref{eqbb3}) and (\ref{eqbbc4}) obey the following symmetry relations
\begin{equation}\label{eqA58}
\rho_{ik}=\rho_{ki},\quad \etaa_{ill}=-\thetaa_{mmi}.
\end{equation}
We reorganise the constitutive relations into a set of explicit expressions for $\m_i$ and $\Theta$. To this end we divide both sides of equation (\ref{eqbbc4})  by $K$, which gives
\begin{eqnarray}\label{eqA59}
&&\Theta=-\zeta_pv_p -\kappa\sigmaa,
\end{eqnarray}
with
\begin{eqnarray}
&&\zeta_p=\third\kappa\thetaa_{mmp},\\
&&\kappa=1/K.
\end{eqnarray}
Substitution   into equation (\ref{eqbb3}) gives
\begin{equation}\label{eqA61}
\m_i={\barrho}_{ip}v_p+\xi_i\sigmaa,
\end{equation}
with
\begin{eqnarray}
&&{\barrho}_{ip}=\rho_{ip}-\ninth\kappa\etaa_{ill}\thetaa_{mmp},\\
&&\xi_i=-\third\kappa\etaa_{ill}.
\end{eqnarray}
Equations (\ref{eqA61}) and (\ref{eqA59}) form a new set of constitutive relations. On account of equation (\ref{eqA58}), the medium parameters in these relations obey the following symmetry relations
\begin{equation}
{\barrho}_{ip}={\barrho}_{pi},\quad \xi_p=\zeta_p.
\end{equation}
Substitution of constitutive relations (\ref{eqA61}) and (\ref{eqA59}) into equations (\ref{eqbb1}) and (\ref{eqbb2}), using $\xi_p=\zeta_p$, 
gives
\begin{eqnarray}
&&{\barrho}_{ip}\partial_tv_p+\xi_i\partial_t\sigmaaa+\partial_i\sigmaaa=\F_i,\label{eqbb98}\\
&&\xi_p\partial_tv_p + \kappa\partial_t\sigmaaa  + \partial_kv_k=q.\label{eqbb99}
\end{eqnarray}
Next, we assume that the wave fields, sources and medium parameters are independent of the $x_2$-coordinate.
Furthermore, we assume 
${\barrho}_{12}={\barrho}_{32}=0$
and $\xi_2=0$.
Then equation  (\ref{eqbb98}) for $i=1,3$ (using ${\barrho}_{13}={\barrho}_{31}$) and equation (\ref{eqbb99})  yield three equations,
describing the propagation of acoustic (AC) waves
 (with wave field quantities $\sigmaaa$,  $v_1$ and $v_3$) in the $(x_1,x_3)$-plane:
\begin{eqnarray}
&&\kappa\partial_t\sigmaaa +\xi_1\partial_tv_1 +\xi_3\partial_tv_3 + \partial_1v_1 + \partial_3v_3=q,\label{eqAC1}\\
&&{\barrho}_{11}\partial_tv_1+{\barrho}_{31}\partial_tv_3+\xi_1\partial_t\sigmaaa+\partial_1\sigmaaa=\F_1,\label{eqAC2}\\
&&{\barrho}_{31}\partial_tv_1+{\barrho}_{33}\partial_tv_3+\xi_3\partial_t\sigmaaa+\partial_3\sigmaaa=\F_3.\label{eqAC3}
\end{eqnarray}

\subsection{Unified scalar wave equation}

The systems of equations for 
transverse-electric waves (TE waves, equations (\ref{eqTE1}) $-$ (\ref{eqTE3})),
transverse-magnetic waves (TM waves, equations (\ref{eqTM1}) $-$ (\ref{eqTM3})),
horizontally polarised shear waves (SH waves, equations (\ref{eqSH1}) $-$ (\ref{eqSH3})) and
acoustic waves (AC waves, equations (\ref{eqAC1}) $-$ (\ref{eqAC3})),
can all be cast in the following form
\begin{eqnarray}
&&\alpha \partial_tP +   (\partial_r+ \gamma_r\partial_t)Q_r =B,\label{eq15agt}\\
&&(\partial_s+ \gamma_s\partial_t)P+\beta_{su}\partial_tQ_u =C_s,\label{eq16agt}
\end{eqnarray}
with $\beta_{su}=\beta_{us}$. Recall that subscripts $r$, $s$ and $u$ only take the values 1 and 3. 
The field quantities, medium parameters and source functions in these equations are given in Table 1
for TE, TM, SH and AC waves. We derive a scalar wave equation for $P$ by eliminating $Q_r$ from equations (\ref{eq15agt}) and (\ref{eq16agt}).
We define the inverse of $\beta_{su}$ via
\begin{eqnarray}
\s_{rs}\beta_{su}=\delta_{ru}.
\end{eqnarray}
Because $\beta_{su}$ is a symmetric $2\times 2$ tensor, the following simple expressions hold for $\s_{rs}$
\begin{eqnarray}
\s_{11}&=&\beta_{33}/\Delta,\label{eqA73a}\\
\s_{13}=\s_{31}&=&-\beta_{31}/\Delta,\label{eqA74a}\\
\s_{33}&=&\beta_{11}/\Delta,\label{eqA75a}
\end{eqnarray}
with 
\begin{eqnarray}
\Delta=\beta_{11}\beta_{33}-\beta_{31}^2.\label{eqA76a}
\end{eqnarray}
Apply $\partial_t$ to both sides of equation (\ref{eq15agt}) and $(\partial_r+\gamma_r\partial_t)\s_{rs}$ to
both sides of equation (\ref{eq16agt})  and subtract the results. Using the fact that the effective medium parameters are time-independent, this gives
\begin{eqnarray}
&& (\partial_r+\gamma_r\partial_t) \s_{rs}(\partial_s+\gamma_s\partial_t)P-\alpha\partial_t^2P=(\partial_r+\gamma_r\partial_t)\s_{rs}C_s-\partial_tB.\label{eq16aghh}
\end{eqnarray}

\section{Decomposition of the reciprocity theorems for non-reciprocal media}\label{AppB}

We derive (1) a unified matrix-vector wave equation for non-reciprocal media, (2) apply decomposition to the operator matrix, and (3) use the symmetry properties
of the decomposed operators to derive reciprocity theorems for decomposed wave fields. 

\subsection{Unified matrix-vector wave equation}

Using the Fourier transform, defined in equation (\ref{eqFT}), we transform equations (\ref{eq15agt}) and (\ref{eq16agt}) to the space-frequency domain, yielding
\begin{eqnarray}
&&-\i\omega\alpha P + (  \partial_r-\i\omega \gamma_r)Q_r =B,\label{eq15ag}\\
&&(\partial_s-\i\omega \gamma_s)P-\i\omega\beta_{su}Q_u =C_s.\label{eq16ag}
\end{eqnarray}
We derive a matrix-vector wave equation of the form
\begin{eqnarray}\label{eq22}
\partial_3\bq =\bA\bq +\bd,
\end{eqnarray}
with  wave vector $\bq =\bq (\bx,\omega)$  and source vector $\bd =\bd (\bx,\omega)$ defined as
\begin{eqnarray}\label{eq23}
\bq =\begin{pmatrix} P \\ Q_3 \end{pmatrix}, 
\quad \bd =\begin{pmatrix} C^o\\ 
B^o \end{pmatrix}
\end{eqnarray}
and  operator matrix $\bA=\bA(\bx,\omega)$ defined as
\begin{eqnarray}\label{eq24}
\bA=\begin{pmatrix}{\cal A}_{11} & {\cal A}_{12} \\ {\cal A}_{21} & {\cal A}_{22} \end{pmatrix}.
\end{eqnarray}
 \rev{To this end,} we separate the derivatives in the $x_3$-direction from the derivatives in the $x_1$-direction in equations (\ref{eq15ag}) and 
  (\ref{eq16ag}), the latter multiplied by $\s_{33}^{-1}\s_{3s}$ on both sides. Hence,
\begin{eqnarray}
&& \partial_3Q_3=\i\omega\alpha P+ \i\omega \gamma_rQ_r  - \partial_1Q_1 +  B,\label{eq74}\\
&& \partial_3P=  - \s_{33}^{-1} ( -\i\omega Q_3-\i\omega\s_{3s}\gamma_sP + \s_{31}\partial_1P - \s_{3s}C_s).\label{eq73}
\end{eqnarray}
$Q_1$ needs to be eliminated from  equation (\ref{eq74}). From equation (\ref{eq16ag}), multiplied on both sides by $\s_{1s}$, we obtain
\begin{eqnarray}\label{eq20}
&& Q_1=\frac{1}{\i\omega}(-\i\omega \s_{1s}\gamma_sP +\s_{1s}\partial_sP-\s_{1s}C_s).
\end{eqnarray}
Substitution of equation (\ref{eq20}) into (\ref{eq74}) gives
\begin{eqnarray}
&& \partial_3Q_3=\i\omega\alpha P+ \i\omega \gamma_3Q_3  -\frac{1}{\i\omega}( \partial_1-\i\omega\gamma_1) (-\i\omega \s_{1s}\gamma_sP+\s_{1s}\partial_sP -\s_{1s}C_s) +  B,
\end{eqnarray}
or, upon substitution of equation (\ref{eq73}) and some reorganization,
\begin{eqnarray}
&& \partial_3Q_3=\Bigl( \i\omega\alpha - \frac{1}{\i\omega}( \partial_1-\i\omega\gamma_1)\bb_1( \partial_1-\i\omega\gamma_1)\Bigr) P\nonumber\\
&&\hspace{1cm} +\Bigl(\i\omega \gamma_3-( \partial_1-\i\omega\gamma_1)\s_{13}\s_{33}^{-1}\Bigr)Q_3 
+B +\frac{1}{\i\omega}(\partial_1-\i\omega\gamma_1) \bb_sC_s,\label{eq77}
\end{eqnarray}
with
\begin{eqnarray}
&&\bb_s=\s_{1s}-\s_{13}\s_{33}^{-1}\s_{3s},
\end{eqnarray}
or, using equations (\ref{eqA73a}) $-$ (\ref{eqA76a}),
\begin{eqnarray}
&&\bb_1=1/\beta_{11},\\
&&\bb_3=0.
\end{eqnarray}
Equations (\ref{eq73}) and (\ref{eq77}) can be cast in the form of the matrix-vector wave equation defined in equations (\ref{eq22}) $-$ (\ref{eq24}), 
with
\begin{eqnarray}
{\cal A}_{11} &=& \i\omega\gamma_3 -  \ss( \partial_1-\i\omega\gamma_1),\label{eq25}  \\ 
{\cal A}_{12} &=&\i\omega\s_{33}^{-1}, \label{eq26}\\ 
{\cal A}_{21} &=& \i\omega\alpha - \frac{1}{\i\omega}( \partial_1-\i\omega\gamma_1)\bb_1( \partial_1-\i\omega\gamma_1),\label{eq27} \\ 
{\cal A}_{22} &=& \i\omega\gamma_3-( \partial_1-\i\omega\gamma_1)\ss,  \label{eq28}\\
C^o&=&
\ss C_1+C_3,\\
B^o&=&B+\frac{1}{\i\omega}(\partial_1-\i\omega\gamma_1) \bb_1C_1,
\end{eqnarray}
with
\begin{eqnarray}
&&\ss=\s_{33}^{-1}\s_{13}=-\beta_{31}/\beta_{11}.
\end{eqnarray}
The notation in the right-hand side of equations (\ref{eq25}) $-$ (\ref{eq28}) should be understood in the sense that differential operators act on all factors to the right of it. For example,
the operator $\partial_1\bb_1\partial_1$ in equation (\ref{eq27}), applied via equation (\ref{eq22}) to the wave field $P$, implies $\partial_1(\bb_1\partial_1P)$.

\subsection{Decomposition of the operator matrix}

We use equation (\ref{eqFTP}) to transform the operator matrix $\bA$ defined in equation (\ref{eq24})  to the slowness domain, 
assuming the medium is laterally invariant at depth level $x_3$. 
The spatial differential operators $\partial_1$ are thus replaced by $\i\omega \p $, hence
\begin{eqnarray}\label{eq24pe}
\bAt(\p ,x_3,\omega)=\begin{pmatrix}
 \i\omega\{\gamma_3-\ss(\p-\gamma_1 )\} &\i\omega\s_{33}^{-1} \\
  \i\omega\s_{33} \q^2 & \i\omega\{\gamma_3-\ss(\p-\gamma_1 )\}
   \end{pmatrix},
\end{eqnarray}
with
\begin{eqnarray}
\q^2=\s_{33}^{-1}\bigl(\alpha - \bb_1(\p-\gamma_1 )^2\bigr).
\end{eqnarray}
The eigenvalue decomposition of $\bAt$ reads
\begin{eqnarray}\label{eqALHL}
\bAt=\bLt\bHt\bLt^{-1}.
\end{eqnarray}
Using the standard approach to find eigenvalues and eigenvectors we obtain
\begin{eqnarray}
\bHt(\p ,x_3,\omega)
&=&\begin{pmatrix}\i\omega\lambda^+ & 0 \\ 0 & -\i\omega\lambda^-\end{pmatrix},\\
\bLt(\p ,x_3,\omega)&=&\frac{1}{\sqrt{2}}\begin{pmatrix}1/\sqrt{\h\q} & 1/\sqrt{\h\q}\\  \sqrt{\h\q} &  -\sqrt{\h\q}   \end{pmatrix},\label{eqAL}\\
\{\bLt(\p ,x_3,\omega)\}^{-1}&=&\frac{1}{\sqrt{2}}\begin{pmatrix}\sqrt{\h\q} & 1/\sqrt{\h\q}\\  \sqrt{\h\q} &  -1/\sqrt{\h\q}   \end{pmatrix},
\end{eqnarray}
where
\begin{eqnarray}
\lambda^\pm&=&\q\pm\{\gamma_3-\ss(\p-\gamma_1)\},\\
\q&=&
\begin{cases}
\sqrt{\s_{33}^{-1}\bigl(\alpha - \bb_1(\p-\gamma_1 )^2\bigr)},\quad\mbox{for}\quad (\p-\gamma_1 )^2\le \frac{\alpha}{\bb_1},  \\
\i \sqrt{\s_{33}^{-1}\bigl(\bb_1(\p-\gamma_1 )^2-\alpha\bigr)},\quad\mbox{for}\quad (\p-\gamma_1 )^2> \frac{\alpha}{\bb_1}.
\end{cases}\label{eqAp3}
\end{eqnarray}
Note that the intervals $(\p-\gamma_1 )^2\le \frac{\alpha}{\bb_1}$ and $(\p-\gamma_1 )^2>\frac{\alpha}{\bb_1}$ in equation (\ref{eqAp3})
 \rev{correspond to} propagating  and  evanescent waves, respectively.

\subsection{Reciprocity theorems for decomposed wave fields}

We derive reciprocity theorems for downgoing and upgoing  \rev{flux-normalized} wave fields, exploiting the symmetry properties of  operator $\bLt$. 
Reciprocity theorems (\ref{eq19g}) and (\ref{eq20g}) can be compactly written as

\begin{eqnarray}\label{eq19gap}
&&\int_{\setdD_0}\{\bq_A^{\a}\}^t\bN\bq_B\dx=\int_{\setdD_A}\{\bq_A^{\a}\}^t\bN\bq_B\dx
\end{eqnarray}
and
\begin{eqnarray}\label{eq20gap}
&&\int_{\setdD_0}\bq_A^\dagger\bK\bq_B\dx=\int_{\setdD_A}\bq_A^\dagger\bK\bq_B\dx,
\end{eqnarray}
with $\bq$ defined in equation (\ref{eq23ag}),
 superscript $t$ denoting transposition, $\dagger$ transposition and complex conjugation,  and matrices $\bN$ and $\bK$  
defined as
\begin{eqnarray}\label{eq4.3a}
{\bN}=\begin{pmatrix} 0 & 1 \\ -1 & 0 \end{pmatrix},
\quad {\bK}=\begin{pmatrix} 0 & 1 \\ 1 & 0 \end{pmatrix}.
\end{eqnarray}
According to equation (\ref{eqcomp}), vector $\bq$ is (for both states) related to vector $\bp$ via $\bq=\bL\bp$, with $\bp$ defined in equation (\ref{eq23ag}).
Here we use this relation and the symmetry properties of composition operator $\bLt$ 
to recast equations (\ref{eq19gap}) and (\ref{eq20gap}) into reciprocity theorems for downgoing and upgoing wave fields.
 
Using the spatial Fourier transform, defined in equation (\ref{eqFTP}), and Parseval's theorem, we first rewrite the integrals in equations (\ref{eq19gap}) and (\ref{eq20gap}) as
\begin{eqnarray}\label{eq19gapp1}
&&\int_{-\infty}^\infty\{\bq_A^{\a}(x_1,x_3,\omega)\}^t\bN\bq_B(x_1,x_3,\omega)\dxx=\\
&&\hspace{1cm}\frac{\omega}{2\pi}\int_{-\infty}^\infty\{\tilde\bq_A^{\a}(-\p ,x_3,\omega)\}^t\bN\tilde\bq_B(\p ,x_3,\omega)\dpp\nonumber
\end{eqnarray}
and
\begin{eqnarray}\label{eq20gapp1}
&&\int_{-\infty}^\infty\{\bq_A(x_1,x_3,\omega)\}^\dagger\bK\bq_B(x_1,x_3,\omega)\dxx=\\
&&\hspace{1cm}\frac{\omega}{2\pi}\int_{-\infty}^\infty\{\tilde\bq_A(\p ,x_3,\omega)\}^\dagger\bK\tilde\bq_B(\p ,x_3,\omega)\dpp,\nonumber
\end{eqnarray}
respectively, where $x_3$ can represent the depth level of $\setdD_0$ or $\setdD_A$. 
Assuming the medium parameters are laterally invariant at $x_3$, the composition operation
$\bq=\bL\bp$ can be rewritten in the slowness domain as
\begin{eqnarray}
\tilde\bq(\p ,x_3,\omega)=\bLt(\p ,x_3,\omega)\tilde\bp(\p ,x_3,\omega),
\end{eqnarray}
with $\bLt(\p ,x_3,\omega)$ defined in equation (\ref{eqAL}). Substituting this in the right-hand sides of equations (\ref{eq19gapp1}) and (\ref{eq20gapp1}) yields
\begin{eqnarray}\label{eq19gapp2}
&&\frac{\omega}{2\pi}\int_{-\infty}^\infty\{\tilde\bq_A^{\a}(-\p ,x_3,\omega)\}^t\bN\tilde\bq_B(\p ,x_3,\omega)\dpp=\\
&&\hspace{1cm}\frac{\omega}{2\pi}\int_{-\infty}^\infty\{\tilde\bp_A^{\a}(-\p ,x_3,\omega)\}^t\{\bLt^{\a}(-\p ,x_3,\omega)\}^t\bN
\bLt(\p ,x_3,\omega)\tilde\bp_B(\p ,x_3,\omega)\dpp\nonumber
\end{eqnarray}
and
\begin{eqnarray}\label{eq20gapp2}
&&\frac{\omega}{2\pi}\int_{-\infty}^\infty\{\tilde\bq_A(\p ,x_3,\omega)\}^\dagger\bK\tilde\bq_B(\p ,x_3,\omega)\dpp=\\
&&\hspace{1cm}\frac{\omega}{2\pi}\int_{-\infty}^\infty\{\tilde\bp_A(\p ,x_3,\omega)\}^\dagger\{\bLt(\p ,x_3,\omega)\}^\dagger\bK
\bLt(\p ,x_3,\omega)\tilde\bp_B(\p ,x_3,\omega)\dpp,\nonumber
\end{eqnarray}
respectively. From the definition of $\bLt(\p ,x_3,\omega)$ in equation (\ref{eqAL}), with $\q$ defined in equation (\ref{eqAp3}), recalling that superscript $\a$ implies that 
$\gamma_r$ is replaced by $-\gamma_r$, we find
\begin{eqnarray}
\{\bLt^{\a}(-\p,x_3,\omega)\}^t\bN\bLt(\p,x_3,\omega)&=&-\bN,\quad\mbox{for}\quad -\infty<\p <\infty,\label{eq120}\\
\{{\bLt}(\p,x_3,\omega)\}^\dagger\bK\bLt(\p,x_3,\omega)&=&\bJ,\quad\mbox{for}\quad (\p-\gamma_1 )^2\le \frac{\alpha}{\bb_1}, \label{eq121}
\end{eqnarray}
with $\bJ$ defined as
\begin{eqnarray}\label{eq4.3all}
\quad {\bJ}=\begin{pmatrix} 1 & 0 \\ 0 & -1 \end{pmatrix}.
\end{eqnarray}
Note that equation (\ref{eq120}) holds for propagating and evanescent waves, whereas 
 equation (\ref{eq121}) holds for propagating waves only.
Substituting equations (\ref{eq120}) and (\ref{eq121}) into equations (\ref{eq19gapp2}) and (\ref{eq20gapp2}) 
and using Parseval's theorem again yields
\begin{eqnarray}\label{eq19gapp4}
&&\int_{-\infty}^\infty\{\bq_A^{\a}(x_1,x_3,\omega)\}^t\bN\bq_B(x_1,x_3,\omega)\dxx=\\
&&\hspace{1cm}-\int_{-\infty}^\infty\{\bp_A^{\a}(x_1,x_3,\omega)\}^t\bN\bp_B(x_1,x_3,\omega)\dxx\nonumber
\end{eqnarray}
and
\begin{eqnarray}\label{eq20gapp4}
&&\int_{-\infty}^\infty\{\bq_A(x_1,x_3,\omega)\}^\dagger\bK\bq_B(x_1,x_3,\omega)\dxx=\\
&&\hspace{1cm}\int_{-\infty}^\infty\{\bp_A(x_1,x_3,\omega)\}^\dagger\bJ\bp_B(x_1,x_3,\omega)\dxx,\nonumber
\end{eqnarray}
respectively. Equation (\ref{eq19gapp4}) is exact, whereas in equation (\ref{eq20gapp4}) evanescent waves are neglected. 
Using these equations at boundaries $\setdD_0$ and $\setdD_A$ in reciprocity theorems   (\ref{eq19gap}) and (\ref{eq20gap})  yields
\begin{eqnarray}\label{eq19gapp5}
&&\int_{\setdD_0}\{\bp_A^{\a}\}^t\bN\bp_B\dx=\int_{\setdD_A}\{\bp_A^{\a}\}^t\bN\bp_B\dx
\end{eqnarray}
and
\begin{eqnarray}\label{eq20gapp5}
&&\int_{\setdD_0}\bp_A^\dagger\bJ\bp_B\dx=\int_{\setdD_A}\bp_A^\dagger\bJ\bp_B\dx,
\end{eqnarray}
respectively. Substituting the expressions for $\bp$ (equation \ref{eq23ag}), $\bN$ (equation \ref{eq4.3a}) and $\bJ$ (equation \ref{eq4.3all}) 
we obtain the reciprocity theorems of equations  (\ref{eq142}) and (\ref{eq143}) for the downgoing and upgoing fields $\u^+$ and $\u^-$.

\newpage
\centerline{{\Large References}}


\end{spacing}

\end{document}